\documentclass[preprint,showpacs]{revtex4}
\usepackage{graphicx}
\usepackage{bm}
\usepackage{color}
\usepackage{amsmath}
\usepackage{natbib}

\begin{document}

\title{The quantum hydrodynamic description of quantum gases with different interactions}

\author{Pavel A. Andreev}
\email{andreevpa@physics.msu.ru}
\affiliation{M. V. Lomonosov Moscow State University, Moscow, Russia.}

 \date{\today}

\begin{abstract}

The method of derivation of equations describing the evolution of the neutral Bose particle system at low temperatures is described. Despite the fact that we consider the neutral particles we account the short-range interaction between particles. As important limiting case we consider the particles in the Bose-Einstein condensate (BEC) state. This method is called the method of quantum hydrodynamics, because natural for of the quantum mechanics rewritten in terms of material fields of observable quantities in three dimensional space is the set of equations, which look like the hydrodynamics equations.
It can be shown that from the quantum hydrodynamics (QHD) equations can be derived macroscopic non-linear Schrodinger equation. Most famous non-linear Schrodinger equation is the Gross-Pitaevskii (GP) equation, which is non-linear Schrodinger equation with nonlinearities of the third degree. Non-linear Schrodinger equation defines the wave function in the medium or order parameter, which is a macroscopic parameter. There are generalizations of the GP equation. New term appears in the GP equation at account of the three-particle interaction. This term contains nonlinearity of the fifth degree. At more detailed, in comparison with the GP equation, account of the two particle interaction we come to the non-local non-linear Schrodinger equation. This equation contains spatial derivatives of the order parameter in the non-linear terms caused by the interaction. Particularly, non-local non-linear Schrodinger equation arises at the consideration of the two-particle interaction up to the third order by the interaction radius. In this terminology the GP equation corresponds to the first order by the interaction radius.
For the BEC of the neutral particles with anisotropic long-range dipole-dipole interaction the generalization of the GP equation was also suggested. Detailed analyses of the applicability conditions shows that this equation valid for the system of dipoles parallel to each other, which do not change their direction, and where the dipole-dipole interaction interferences � motion of particles.
All described non-linear Schrodinger equation can be derived from the corresponding, and in some cases more general, QHD equations. And, for all described cases and types of interaction, the equations of the QHD can be derived directly from the microscopic many-particle Schrodinger equation.
This chapter is dedicated to the description of the method of the QHD. We show the method of derivation of the hydrodynamics equation and present the method of derivation of the corresponding non-linear Schrodinger equation from the QHD equations. During derivation we admit advantages of the QHD method.
The QHD equations contain information about thermal motion of the particles, when we consider the BEC evolution we have to neglect by the contribution of the thermal motion. In the chapter we illustrate contribution of the temperature on the Bose particle dynamics at the example of the system of Bose particles with the three particle interaction in the first order by the interaction radius approximation.
Finally, we discuss new properties of the dispersion dependency of the eigenwaves in the BEC obtained recently by means of the QHD.

\end{abstract}

\maketitle


\section{Introduction}

The model of a slightly non-ideal Bose gas at almost zero temperature is a fundamental model for the BEC description giving Bose-type energy spectrum suggested by N.N. Bogoliubov \cite{Shirkov UFN 09}, \cite{Landau Vol 9}. The non-linear Schrodinger equation for the wave function of BEC was suggested later, which is called the Gross-Pitaevskii equation \cite{Landau Vol 9}, \cite{L.P.Pitaevskii RMP 99}. It is a great tool for inhomogeneous BEC studying. This equation contains nonlinearity of third degree. The Gross-Pitaevskii equation was generalized for the BEC of particles with three-particle interaction \cite{Kovalev FNT 76}, \cite{Barashenkov PLA 88} (see also more recent papers \cite{Abdullaev PRA 01}-\cite{Jack PRL 02}). In this case a nonlinearity of fifth degree is added to describe three-particle interaction. A generalization of the Gross-Pitaevskii equation for dipolar BEC was also suggested \cite{Goral PRA 00}-\cite{Fischer PRA 06R}. It is well-known that the Gross-Pitaevskii equation might be presented in hydrodynamical form, analogously to the hydrodynamical representation of the one-particle Schrodinger equation.

We consider evolution of the BEC from point of view of the quantum hydrodynamic method. Initially the QHD method was developed for quantum plasma \cite{MaksimovTMP 1999}-\cite{Andreev PRB 11}. Quantum hydrodynamics of ultracold quantum gases was obtained later \cite{Andreev PRA08}. The QHD method allows us to study evolution of spinning particles \cite{MaksimovTMP 2001}, \cite{Marklund PRL 07} and particles having electric dipole moment \cite{Andreev PRB 11}. Method of the QHD allows us to derive equations of the quantum observable directly from many-particle Schrodinger equation. These equations arise as a chain of equations, and the chain has to be truncated. A way of truncation depends on the kind of interparticle interaction and a set of physical process we want to consider. The approximation of the self-consistent field approximation gives us possibility to get a closed set of equation for particles with a long-range interaction (we use it for modeling of the dipole-dipole interaction). A short-range interaction requires another treatment. The procedure was developed in Ref. \cite{Andreev PRA08}, and it has no specific name. We will present detailed description of this method in this chapter below.

\section{General structure of the QHD}

This chapter is dedicated to description of the quantum hydrodynamics method in application to the dynamics of Bose-Einstein condensate (BEC) of neutral atoms \cite{Andreev PRA08}. Certainly, we will describe relation between the quantum hydrodynamics and the Gross-Pitaevskii equation, which is usually used for description of the BEC evolution, but we start from basic ideas lay behind the quantum hydrodynamics, which actually follow from basic principles of quantum mechanics. One of the main goals we want to show is that the quantum hydrodynamics is a natural way of description of quantum many-particle phenomenon, along with the second quantization.
Then we study dynamics of a quantum physical quantity we have deal with the operator of this variable. Knowledge of wave function allows us to calculate quantum mechanical average of the physical quantity \cite{Landau Vol 3}
$$<L>=\int \psi^{*}\hat{L}\psi dR,$$
then the wave function satisfy to the Schrodinger equation
\begin{equation}\label{ch Ham SRI}\hat{H}=\sum_{i}\frac{1}{2m_{i}}\hat{p}^{\alpha}_{i}\hat{p}_{\alpha i}+\sum_{i}V_{ext}(\textbf{r}_{i},t)+\frac{1}{2}\sum_{i,j,i\neq j}U(\mid \textbf{r}_{i}-\textbf{r}_{j}\mid) ,\end{equation}
where $\hat{p}^{\alpha}_{i}=-\imath\hbar\nabla^{\alpha}_{i}$ is the momentum operator of the i-th particle, $m_{i}$ is the mass of the i-th particle, and $U_{ij}=U(\mid\textbf{r}_{i}-\textbf{r}_{j}\mid)$ is the interaction potential.

If we study a many-particle system we should choose useful physical quantities and find operators for these quantities. Development of the QHD shows that the first of the useful variables is the particle concentration, and it is easy to find it's operator in the coordinate representation. We just need to take the microscopic concentration of particles \cite{Klimontovich book}, \cite{Weinberg book}
$$\hat{n}=\sum_{i=1}^{N}\delta(\textbf{r}-\textbf{r}_{i}).$$
Using this operator we get the quantum mechanical definition of the particle concentration
\begin{equation}\label{ch concentration}n(\textbf{r},t)=\int
 dR\sum_{i}\delta(\textbf{r}-\textbf{r}_{i})\psi^{+}(R,t)\psi(R,t),\end{equation}
where $dR=\prod_{i=1}^{N}d\textbf{r}_{i}.$

We can use the Schrodinger equation to study evolution of the concentration. For this purpose we differentiate the definition of concentration and obtain the continuity equation
\begin{equation}\label{ch continuity eq}\partial_{t}n+\nabla \textbf{j}=0,\end{equation}
where
\begin{equation}\label{ch current}
j^{\alpha}(\textbf{r},t)=\int dR\sum_{i}\delta(\textbf{r}-\textbf{r}_{i})\frac{1}{2m_{i}}\biggl((\hat{p}^{\alpha}_{i}\psi)^{+}(R,t)\psi(R,t)+\psi^{+}(R,t)(\hat{p}^{\alpha}_{i}\psi)(R,t)\biggr)\end{equation}
is the momentum density.

Analogous derivation, but for one-particle Schrodinger equation, is presented in many books in quantum physics.
In one particle case our definition (\ref{ch concentration}) gives well-known formula for density of probability $\psi^{*}\psi$.

Our definition (\ref{ch concentration}) has one more important feature. Many-particle wave function $\psi(R,t)$ is defined in 3N dimensional configuration space, whereas many-particle phenomenon realize in three dimensional physical space. Thus, if we want to have a description method presenting this evolution in natural way we should make projection of 3N dimensional evolution of psi function on three dimensional physical space. However, we do not need to solve this problem since we have solved this problem by using of the definition (\ref{ch concentration}) for the particle concentration.

Having definition of the particle concentration we can derive the set of the QHD equation from the many-particle Schrodinger equation. We have already derived the continuity equation (\ref{ch continuity eq}), this equation contains new physical quantity. It is the particles current or the momentum density. To continue derivation of the QHD set of equations we need to derive an evolution equation of the momentum density, which called the momentum balance equation or the Euler equation. For this purpose we need to differentiate the momentum density with respect to time and use the Schrodinger equation for time derivatives of the wave function. In the result we find
\begin{equation}\label{ch Euler general form}\partial_{t}j^{\alpha}+\partial^{\beta}\Pi^{\alpha\beta}=F^{\alpha},\end{equation}
where $\Pi^{\alpha\beta}$ is the momentum flux tensor, and $F^{\alpha}$ is the force density.

We have found general form of the quantum Euler equation. The momentum flux tensor describes kinetic and kinematic properties. Kinematic properties of spinning and spinless particles are different. In this chapter we will consider spinless particles and we will consider general properties and explicit form of the momentum flux tensor. Whereas, the force density is determined by an explicit form of the interaction in the basic Hamiltonian. We will consider two- and three-particle short range interaction and the long-range dipole-dipole interaction of particles having electric dipole moment. For simplicity we start derivation of the QHD equation for particles with the two-particle short-range interaction. Nevertheless it is enough to derive the Gross-Pitaevskii equation.

We have described general picture. Now we present derivation of the explicit form of the momentum flux tensor and it's connection with the velocity field.

Explicit form of the momentum flux tensor arising in the Euler equation (\ref{ch Euler general form}) is
$$\Pi^{\alpha\beta}(\textbf{r},t)=\int dR\sum_{i}\delta(\textbf{r}-\textbf{r}_{i})\frac{1}{4m_{i}}\biggl(\psi^{+}(R,t)(\hat{p}^{\alpha}_{i}\hat{p}^{\beta}_{i}\psi)(R,t)$$
\begin{equation}\label{ch Pi general form}+(\hat{p}^{\alpha}_{i}\psi)^{+}(R,t)(\hat{p}^{\beta}_{i}\psi)(R,t)+c.c.\biggr). \end{equation}

To present the velocity field in the QHD equations we need to present the many-particle wave function via two real functions as
\begin{equation}\label{ch wave func via exp}\psi(R,t)=a(R,t) exp\biggl(\frac{\imath S(R,t)}{\hbar}\biggr).\end{equation}
Average of square of the amplitude of the wave function $a^{2}$ gives the particle concentration, then the gradient of the phase of the wave function on coordinate of i-th particle gives velocity of i-th particle
\begin{equation}\label{ch velocity of i th part}\textbf{v}_{i}(R,t)=\frac{1}{m_{i}}\nabla_{i}S(R,t),\end{equation}
which actually depends on coordinate of all particles in the system.

Traditionally hydrodynamics equations contain velocity field $\textbf{v}(\textbf{r},t)$, which is velocity of the local center of mass, instead of the momentum density. These quantity simply relate
\begin{equation}\label{ch velocity field definition}\textbf{j}(\textbf{r},t)=n(\textbf{r},t)\textbf{v}(\textbf{r},t).\end{equation}
Having the velocity field and the velocities of each particle we can introduce the difference between velocity of each particle and the velocity field
$\textbf{u}_{i}(\textbf{r},R,t)=\textbf{v}_{i}(R,t)-\textbf{v}(\textbf{r},t)$,
what gives the thermal velocities.

Putting the presentation (\ref{ch wave func via exp}) of the wave function into the momentum flux tensor definition (\ref{ch Pi general form}) and expressing result via both the velocity field and the thermal velocities we get the momentum flux tensor in the following form
\begin{equation}\label{ch Pi representation} \Pi^{\alpha\beta}=nv^{\alpha}v^{\beta}+p^{\alpha\beta}+T^{\alpha\beta},\end{equation}
where $p^{\alpha\beta}$ is the thermal pressure, and $T^{\alpha\beta}$ is the quantum Bohm potential, which is the quantum pressure caused by the quantum nature of particles as the de-Broule wave.

Explicit form of the thermal pressure is
\begin{equation}\label{ch pressure} p^{\alpha\beta}(\textbf{r},t)=\int dR\sum_{i=1}^{N}\delta(\textbf{r}-\textbf{r}_{i})a^{2}(R,t)m_{i}u^{\alpha}_{i}u^{\beta}_{i} ,\end{equation}
as we may expect the  thermal pressure depends on the square of the thermal velocities.
General explicit form of the quantum Bohm potential arises as
\begin{equation}\label{ch Bohm tensor general} T^{\alpha\beta}(\textbf{r},t)=-\frac{\hbar^{2}}{2m}\int dR\sum_{i=1}^{N}\delta(\textbf{r}-\textbf{r}_{i})a^{2}(R,t)\frac{\partial^{2}\ln a}{\partial x_{\alpha i}\partial x_{\beta i}}.\end{equation}
This general formula does not allow us to connect it with the particles concentration and the velocity field to get closed set  of the QHD equations.
In order to obtain the closed set of the QHD equation we consider this term in approximation of independent particles. In this case we find that the quantum Bohm potential in the following form
\begin{equation}\label{ch Bom tensor one part}
T^{\alpha\beta}(\textbf{r},t)=-\frac{\hbar^{2}}{4m}\biggl(\partial^{\alpha}\partial^{\beta}n(\textbf{r},t)-\frac{1}{n(\textbf{r},t)}(\partial^{\alpha}n(\textbf{r},t))(\partial^{\beta}n(\textbf{r},t))\biggr),\end{equation}
where we find that the quantum Bohm potential expressed via the spatial derivatives of the particle concentration.
Divergence of this tensor has well-known in literature form
\begin{equation} \label{ch Bom diverg of it}\partial_{\beta}T^{\alpha\beta}(\textbf{r},t)=-\frac{\hbar^{2}}{2m}n(\textbf{r},t)\partial_{\alpha}\frac{\triangle\sqrt{n(\textbf{r},t)}}{\sqrt{n(\textbf{r},t)}} .\end{equation}

Now we have represented the hydrodynamical quantities via velocity field and got explicit form of the quantum Bohm potential we can rewrite the set of QHD equation
\begin{equation}\label{ch continuity via vel}
\partial_{t}n(\textbf{r},t)+\nabla(n(\textbf{r},t)\textbf{v}(\textbf{r},t))=0,\end{equation}
and
\begin{equation}\label{ch eiler via vel} mn(\textbf{r},t)(\partial_{t}+\textbf{v}\nabla)v^{\alpha}(\textbf{r},t)+\partial_{\beta}(p^{\alpha\beta}(\textbf{r},t)+T^{\alpha\beta}(\textbf{r},t))=F^{\alpha} (\textbf{r},t).\end{equation}
This equations have traditional form. Now we should consider the right-hand side of the Euler equation (\ref{ch eiler via vel and stress tensor}), where we have the force field density. We have deal with the system of neutral particles interacting by means of the short-range interaction potential. We can use this main properties of the interaction to construct a general hydrodynamical theory for such system.

\section{short range interaction in QHD}

Let's start studying of the explicit form of the force density.
We have not presented explicit form of the force density obtained in equation (\ref{ch Euler general form}). The force density consists of two parts, which are action of an external field and an interparticle interaction, thus we can write
\begin{equation} \label{ch } \textbf{F}=-n\nabla V+\textbf{F}_{int}\end{equation}
and  the second term, which is under our main interest, has following form
\begin{equation} \label{ch F int explicit} \textbf{F}_{int}=-\int dR\sum_{i,j\neq i}\delta(\textbf{r}-\textbf{r}_{i})(\nabla_{i}U(\textbf{r}_{ij}))\psi^{+}(R,t)\psi(R,t) .\end{equation}
There is no difference between i-th particle and j-th particle, thus we can rewrite it as
\begin{equation} \label{ch F int explicit j} \textbf{F}_{int}=-\int dR\sum_{i,j\neq i}\delta(\textbf{r}-\textbf{r}_{j})(\nabla_{j}U(\textbf{r}_{ij}))\psi^{+}(R,t)\psi(R,t) .\end{equation}
Using symmetry of the wave function to permutation of arguments $\textbf{r}_{i}$ and $\textbf{r}_{j}$, and the fact that $\nabla_{j}U(\textbf{r}_{ij})=-\nabla_{i}U(\textbf{r}_{ij})$ we can rewrite formula (\ref{ch F int explicit j}) as
\begin{equation} \label{ch F int explicit j transformed} \textbf{F}_{int}=\int dR\sum_{i,j\neq i}\delta(\textbf{r}-\textbf{r}_{j})(\nabla_{i}U(\textbf{r}_{ij}))\psi^{+}(R,t)\psi(R,t) .\end{equation}
Keeping in mind that formulas (\ref{ch F int explicit}) and (\ref{ch F int explicit j transformed}) are different forms of the same quantity $\textbf{F}_{int}$. Thus, we can present $\textbf{F}_{int}$ as half of the sum of (\ref{ch F int explicit}) and (\ref{ch F int explicit j transformed}), and consequently we find
\begin{equation} \label{ch F int explicit j final} \textbf{F}_{int}=-\frac{1}{2}\int dR\sum_{i,j\neq i}(\delta(\textbf{r}-\textbf{r}_{i})-\delta(\textbf{r}-\textbf{r}_{j}))(\nabla_{i}U(\textbf{r}_{ij}))\psi^{+}(R,t)\psi(R,t) ,\end{equation}
This formula is very useful for further transformation and let us proceed in (\ref{ch F int explicit j final}) to variables of the center of gravity and variables of the relative distance of the particles defined as
\begin{equation} \label{ch def of R and r} \begin{array}{ccc} \textbf{R}_{ij}=\frac{1}{2}(\textbf{r}_{i}+\textbf{r}_{j}) ,& \textbf{r}_{ij}=\textbf{r}_{i}-\textbf{r}_{j} \end{array}.\end{equation}
Putting new variables in the force density $\textbf{F}_{int}$ we obtain
$$\textbf{F}_{int}=-\frac{1}{2}\int dR\sum_{i,j\neq i}\biggl(\delta(\textbf{r}-\textbf{R}_{ij}-\frac{1}{2}\textbf{r}_{ij})-\delta(\textbf{r}-\textbf{R}_{ij}+\frac{1}{2}\textbf{r}_{ij})\biggr)(\nabla_{i}U(\textbf{r}_{ij}))$$
\begin{equation} \label{ch F int explicit j final new var} \psi^{+}(R_{N-2},\textbf{R}_{ij}+\frac{1}{2}\textbf{r}_{ij},\textbf{R}_{ij}-\frac{1}{2}\textbf{r}_{ij},t)\psi(R_{N-2},\textbf{R}_{ij}+\frac{1}{2}\textbf{r}_{ij},\textbf{R}_{ij}-\frac{1}{2}\textbf{r}_{ij},t) ,\end{equation}
where $\textbf{R}_{ij}+\frac{1}{2}\textbf{r}_{ij}$ stands for $i$-th argument of the wave function, $\textbf{R}_{ij}-\frac{1}{2}\textbf{r}_{ij}$ is the same for $j$-th argument, and $R_{N-2}$ is the set of rest coordinates.

Since the interaction forces between the particles rapidly descend at distances of the order of the interaction radius, small $|r^{\alpha}_{ij}|$ give the main contribution to integral in (\ref{ch F int explicit j final}). Therefore, in expression (\ref{ch F int explicit j final}), we can replace the multipliers at the interaction potential by their expansion in series by $\textbf{r}_{ij}$. It is worthwhile to admit that we should make expansion of three functions, they are the difference between $\delta$ functions, $i$-th and $j$-th arguments in $\psi^{*}(R,t)$ and $\psi(R,t)$. We have reached the conclusion that the density of the interaction force for bosons with a short-range interaction potential can be represented in the form of divergence of the tensor field $\partial_{\beta}\sigma^{\alpha\beta}(\textbf{r},t)$. Here, $\sigma^{\alpha\beta}(\textbf{r},t)$ is the quantum stress tensor conditioned by the occurrence of interparticle interaction. Divergence of this tensor is represented by the formula
$$F^{\alpha}_{int}=-\partial^{\beta}\sigma^{\alpha\beta}(\textbf{r},t)=\frac{1}{2}\int dR\sum_{i,j\neq i}(\partial_{i}^{\alpha}U(\textbf{r}_{ij}))\partial^{\beta}_{\textbf{r}}\biggl(r^{\beta}_{ij}+\frac{1}{24}r^{\beta}_{ij}r^{\gamma}_{ij}r^{\delta}_{ij}\biggr)\delta(\textbf{r}-\textbf{R}_{ij})$$
$$\times\Biggl(\psi^{+}\psi+\frac{1}{2}r^{\mu}_{ij}\biggl(\psi^{+}(\partial^{\mu}_{\textbf{R}_{+}}-\partial^{\mu}_{\textbf{R}_{-}})\psi+c.c.\biggr)$$
$$+\frac{1}{4}r^{\mu}_{ij}r^{\nu}_{ij}\biggl((\psi^{+}(\partial^{\mu}_{\textbf{R}_{+}}\partial^{\nu}_{\textbf{R}_{+}}-2\partial^{\mu}_{\textbf{R}_{+}}\partial^{\nu}_{\textbf{R}_{-}}+\partial^{\mu}_{\textbf{R}_{-}}\partial^{\nu}_{\textbf{R}_{-}})\psi+c.c.)$$
\begin{equation} \label{ch sigma shorted series}+(\partial^{\mu}_{\textbf{R}_{+}}-\partial^{\mu}_{\textbf{R}_{-}})\psi^{+}(\partial^{\nu}_{\textbf{R}_{+}}-\partial^{\nu}_{\textbf{R}_{-}})\psi\biggr)\Biggr),\end{equation}
where $\textbf{R}_{ij}$ stands for $i$-th and $j$-th arguments, $\partial_{\textbf{R}_{+}}$ and $\partial_{\textbf{R}_{-}}$ are derivatives on $i$-th and $j$-th arguments correspondingly. In this expansion we have included three first term of the series.

This expansion is valid due to the fact that an interaction radius is small. The first (second, third) term of this expansion is proportional to the interparticle distance limited by the interaction radius in the first (second, third) degree. Therefore we can call this expansion as expansion in series on the interaction radius. Consequently, the first term in formula (\ref{ch sigma shorted series}) arises in the first order by the interaction radius (FOIR).

We follow the basic paper \cite{Andreev PRA08}, where the QHD method was developed up to the third order by the interaction radius (TOIR).

Using notion of the quantum stress tensor we can rewrite the Euler equation (\ref{ch eiler via vel}) as
\begin{equation}\label{ch eiler via vel and stress tensor} mn(\textbf{r},t)(\partial_{t}+\textbf{v}\nabla)v^{\alpha}(\textbf{r},t)+\partial_{\beta}\biggl(p^{\alpha\beta}(\textbf{r},t)+\sigma^{\alpha\beta}(\textbf{r},t)+T^{\alpha\beta}(\textbf{r},t)\biggr)
=-\frac{1}{m}n(\textbf{r},t)\nabla^{\alpha}V_{ext}(\textbf{r}).\end{equation}

In the classic hydrodynamics of neutral particles a generalized
Euler equation \cite{Landau 6} giving the Navier-Stokes equation
at expansion of the stress tensor in the Taylor series on the
spatial derivatives of the velocity field, restricted by linear
terms only has form
\begin{equation} \label{ch Euler generalized LL6} mn\partial_{t}v^{\alpha}+(\textbf{v}\nabla) v^{\alpha}+\partial^{\beta}P^{\alpha\beta}=0.\end{equation}
In contrast to the general classic Euler equation our derivation gives explicit inner structure of $P^{\alpha\beta}$, that leads to more careful understanding of different physical mechanisms contribution.

\section{quantum hydrodynamics of bosons in the first order by the interaction radius}

We have obtained general form of the quantum stress tensor presented by formula (\ref{ch sigma shorted series}). In this section we expand further transformation of the first term in formula (\ref{ch sigma shorted series}). Let's start transformations with rewriting of the first term in expansion of the quantum stress tensor.
\begin{equation} \label{ch sigma in 1 or} \sigma^{\alpha\beta}(\textbf{r},t)=-\frac{1}{2}\int dR\sum_{i,j.i\neq j}\delta(\textbf{r}-\textbf{R}_{ij})\frac{r^{\alpha}_{ij}r^{\beta}_{ij}}{\mid\textbf{r}_{ij}\mid}\frac{\partial U(\textbf{r}_{ij})}{\partial\mid\textbf{r}_{ij}\mid}\psi^{+}(R,t)\psi(R,t).\end{equation}
In the integral (\ref{ch sigma in 1 or}) $dR$ can be rewritten as $dR_{N-2}d\textbf{r}_{i}d\textbf{r}_{j}$ explicitly
distinguish integration on coordinates of $i$-th and $j$-th particles. Using variables of the center of gravity and the relative distance of $i$-th and $j$-th particles we write $dR_{N-2}d\textbf{r}_{i}d\textbf{r}_{j}=dR_{N-2}d\textbf{r}_{ij}d\textbf{R}_{ij}$. It is also important to admit that both $i$-th and $j$-th arguments of the wave function in formula (\ref{ch sigma in 1 or}) are equal to $R_{ij}$, after the expansion in a series on $\textbf{r}_{ij}$. Now we can see that integration on $r_{ij}$ and $R_{ij}$ separates, and we can rewrite formula (\ref{ch sigma in 1 or}) in the following form
\begin{equation}\label{ch sigma in 1 or hz n2}\sigma^{\alpha\beta}(\textbf{r},t)=-\frac{1}{2}Tr(n_{2}(\textbf{r},\textbf{r}',t))\int
d\textbf{r}\frac{r^{\alpha}r^{\beta}}{r}\frac{\partial U(r)}{\partial r} ,\end{equation}
where
$$Tr f(\textbf{r},\textbf{r}')=f(\textbf{r},\textbf{r}),$$
and we have also used notion of the two-particle concentration, which general definition is
\begin{equation}\label{ch two-part conc def}n_{2}(\textbf{r},\textbf{r}',t)=\int dR\sum_{i,j\neq i}\delta(\textbf{r}-\textbf{r}_{i})\delta(\textbf{r}'-\textbf{r}_{j})\psi^{*}(R,t)\psi(R,t). \end{equation}
This definition can be rewritten in more useful form
\begin{equation}\label{ch two-part occup number}n_2(\textbf{r},\textbf{r}',t)=N(N-1)
\int dR_{N-2}\langle n_1,n_2,\ldots |\textbf{r},\textbf{r}',R_{N-2},t\rangle \langle\textbf{r},\textbf{r}',R_{N-2},t |n_1,n_2,\ldots\rangle ,\end{equation}
where $\langle n_1,n_2,\ldots |\textbf{r},\textbf{r}',R_{N-2},t\rangle$ is the N-particle wave function in representation of the occupation numbers, and $dR_{N-2}=\displaystyle\prod\limits_{k=3}^{N}d\textbf{r}_k$.

For further transformation we need to extract evolution of particles related to arguments $\textbf{r}$ and $\textbf{r}'$. For this purpose, we consider expansion of the wave function
$\langle\textbf{r},\textbf{r}',R_{N-2},t |n_1,n_2\ldots\rangle$
\cite{Shveber}. In the case of bosons, making expansion on one of arguments, we find
$$\langle \textbf{r}, \textbf{r}', R_{N-2},t |n_1, n_2 \ldots
\rangle=\sum_f \sqrt{\frac{n_f}{N}} \: \langle\textbf{r},t | f\rangle \:
  \langle \textbf{r}', R_{N-2},t |n_1, \ldots (n_f-1),\ldots \rangle,$$
where we have that particle in  an
arbitrary quantum state $f$ gives dependence on the coordinate $\textbf{r}$, and all particles
alternately make contribution in $\langle\textbf{r},\textbf{r}',R_{N-2},t |n_1,n_2\ldots\rangle$ via $\: \langle\textbf{r},t | f\rangle \:$ due to summation on all states.

Making the second expansion of the wave function, including symmetry of the wave function due to permutation of arguments, we obtain
$$\langle \textbf{r}, \textbf{r}', R_{N-2},t |n_1, n_2 \ldots
\rangle$$
$$=\sum_f \sum_{f', {f'\ne f}}\sqrt{\frac{n_f}{N}} \sqrt{\frac{n_{f'}}{N-1}} \:
 \langle\textbf{r},t | f\rangle \: \langle\textbf{r}',t | f'\rangle\times \langle R_{N-2},t |n_1, \ldots (n_{f'}-1),\ldots (n_f-1), \ldots \rangle$$
\begin{equation}\label{ch second expansion of wave func}+\sum_f \sqrt{\frac{n_f(n_f-1)}{N(N-1)}} \:
 \langle\textbf{r},t | f\rangle \: \langle\textbf{r}',t | f\rangle
 \: \langle R_{N-2},t |n_1, \ldots (n_f-2), \ldots \rangle.\end{equation}
where $\langle \textbf{r},t|f\rangle =\varphi_f(\textbf{r},t)$ are the single-particle wave functions. Formula (\ref{ch second expansion of wave func}) consists of two terms. The first term describes contribution of two particles from different quantum states. The second term gives contribution of two particles being in one quantum states, that corresponds to the fact that several Bose particles can exist in a one quantum state.

Calculation of the two-particle concentration requires integration of the product of two wave functions in formula (\ref{ch two-part occup number})
$$\langle n_1, \ldots (n_{f'}-1), \ldots (n_f-1), \ldots |
  n_1, \ldots (n_{q'}-1),\ldots (n_q-1),\ldots \rangle$$
\begin{equation}\label{ch product of wf}=\delta (f-q) \delta (f'-q')+\delta
(f-q')\delta (f'-q),\end{equation}
and
\begin{equation}\langle n_1, \ldots (n_f-2),\ldots |
n_1, \ldots (n_q-2),\ldots \rangle=\delta (f-q).\end{equation}
Formula (\ref{ch product of wf}) explicitly reveals symmetry of bosons wave function. The second term gives exchange term, and, consequently, contribution of the exchange interaction. Using these formulas, after some calculations, we find following result for the two particle concentration
\begin{equation}\label{ch n2 long r}
n_2(\textbf{r},\textbf{r}',t)=n(\textbf{r},t)n(\textbf{r}',t)+|\rho(\textbf{r},\textbf{r}',t)|^{2}+\sum_{g}n_{g}(n_{g}-1)|\varphi_{g}(\textbf{r},t)|^{2}|\varphi_{g}(\textbf{r}',t)|^{2},\end{equation}
where $n_{g}$ is a number of particles in the quantum state $\varphi_{g}$, with a set of quantum numbers $g$,
\begin{equation}\label{ch nvarphi}
n(\textbf{r},t)=\sum_{g}n_{g}\varphi_{g}^{*}(\textbf{r},t)\varphi_{g}(\textbf{r},t)\end{equation}
is the particle concentration in terms of the arbitrary single-particle wave functions $\varphi_{g}(\textbf{r},t)$,
\begin{equation}\label{ch rhovarphi}\rho(\textbf{r},\textbf{r}',t)=\sum_{g}n_{g}\varphi_{g}^{*}(\textbf{r},t)\varphi_{g}(\textbf{r}',t)\end{equation}
is the macroscopic density matrix.

Putting (\ref{ch n2 long r}) in formula (\ref{ch sigma in 1 or hz n2}) for the quantum stress tensor we obtain
\begin{equation}\label{ch sigma boz}\sigma^{\alpha\beta}(\textbf{r},t)=-\frac{1}{2}\Upsilon\delta^{\alpha\beta}\biggl(2n^{2}(\textbf{r},t)+\wp_{B}(\textbf{r},t)\biggr),\end{equation}
where
$$\wp_{B}(\textbf{r},t)=\sum_{g}n_{g}(n_{g}-1)(|\varphi_{g}(\textbf{r},t)|^{2})^{2},$$
and
\begin{equation}\label{ch Upsilon} \Upsilon=\frac{4\pi}{3}\int
dr(r)^{3}\frac{\partial U(r)}{\partial r}\end{equation}
is the interaction constant appearing from the integral on the relative distance of the particles. Formula (\ref{ch Upsilon}) arises for the spherically symmetric interaction potential. In this chapter we have been developing the theory for neutral particles in the Bose-Einstein condensate state. One of the powerful and famous methods of this studying is the Gross-Pitaevskii equation, which we will derive below from presenting method. Integrating by parts in formula (\ref{ch Upsilon}) and assuming that the potential satisfies the condition that the
quantity $r^{3}U(r)$ tends to zero as $r$ tends to zero and infinity,
we obtain
\begin{equation}\label{g}\Upsilon=-\int d\textbf{r}U(r),\end{equation}
that corresponds to the Gross-Pitaevskii result.

To investigate solitons in the BEC, we use the set of the QHD equations up to the TOIR approximation ~\cite{Andreev PRA08}. The calculation of the first member in a quantum stress tensor that corresponds to the GP equation is made in ~\cite{Andreev PRA08} under the condition that the particles do not interact. A more complete investigation into the conditions of the GP equation derivation from the MPSE shows that the GP equation appears in the first order by the interaction radius (FOIR), if the particles are in an arbitrary state that can be simulated by a single-particle wave function. Such a state can particularly appears as a result of strong interaction between the particles that takes place in the quantum fluids.

$$\wp_{B}(\textbf{r},t)=\sum_{g}n_{g}(n_{g}-1)|\varphi_{g}(\textbf{r},t)|^{4}$$
$$=\sum_{g}n_{g}^{2}|\varphi_{g}(\textbf{r},t)|^{4}=N^{2}|\varphi_{g_{0}}(\textbf{r},t)|^{4}$$
\begin{equation}\label{ch corr.n2}=(N|\varphi_{g_{0}}(\textbf{r},t)|^{2})^{2}=n_{B}^{2}(\textbf{r},t),\end{equation}
Finally, the quantum stress tensor in the BEC state has form
\begin{equation}\label{ch sigma boz only BEC} \sigma_{BEC}^{\alpha\beta}(\textbf{r},t)=-\frac{1}{2} \Upsilon\delta^{\alpha\beta} n^{2}_{B}(\textbf{r},t).\end{equation}
We can see that the quantum stress tensor depends on the constant of interaction $\Upsilon$ and square of the particles concentration.

We have been calculating the quantum stress tensor, which describes interparticle interaction in the Euler equation. We have found $\sigma_{BEC}^{\alpha\beta}$, thus we can put in the Euler equation. Superfluid motion is the eddy-free motion. This means that $\textbf{v}=\nabla\varphi$. For the isotropic kinetic pressure $p^{\alpha\beta}=p \delta^{\alpha\beta}$ under barotropicity condition we introduce the chemical potential $\mu(\textbf{r},t)$ as
\begin{equation}\label{ch mu def}\nabla\mu(\textbf{r},t)=\frac{\nabla p(\textbf{r},t)}{mn(\textbf{r},t)}.\end{equation}
This designation corresponds to the Gross-Pitaevskii equation. However, in the thermodynamics and hydrodynamics $\nabla p(\textbf{r},t)/mn(\textbf{r},t)$ is equal to the gradient of the enthalpy, at the barotropicity condition.

In the result we come to the Euler equation in following form
$$
m\partial_{t}v^{\alpha}(\textbf{r},t)+\frac{1}{2}m\partial^{\alpha}v^{2}(\textbf{r},t)+m\partial^{\alpha}\mu(\textbf{r},t)$$
\begin{equation}\label{ch eiler final}-\frac{\hbar^{2}}{2m}\partial^{\alpha}\frac{\triangle\sqrt{n(\textbf{r},t)}}{\sqrt{n(\textbf{r},t)}}-\Upsilon\partial^{\alpha}n(\textbf{r},t)=-\partial^{\alpha}V_{ext}(\textbf{r},t)\end{equation}
We have obtained final form of the Euler equation. At description of the BEC dynamics it is enough to use the couple of the continuity (\ref{ch continuity via vel}) and Euler (\ref{ch eiler final}) equations, which have been obtained. As it was mentioned the Gross-Pitaevskii equation is usually used for the BEC description. The Gross-Pitaevskii equation is the non-linear Schrodinger equation and it might be easily transformed in the couple of equations having hydrodynamical form. These equations coincide with equations obtained above and now we have the inverse problem. We need to derive the non-linear Schrodinger equation from the set of the QHD equations to present consisted description of the BEC.

As the first step of the non-linear Schrodinger equation derivation we show existence of the Cauchy-Lagrangian integral for the Euler equation (\ref{ch eiler final}). Using notion of the potential of the velocity field we can rewrite equation (\ref{ch eiler final}) as the gradient of the scalar function, which is the Cauchy-Lagrangian integral
\begin{equation}\label{ch int Koshy}\partial_{t}\phi(\textbf{r},t)+\frac{1}{2}(\nabla\phi)^{2}(\textbf{r},t)+\mu(\textbf{r},t)-\frac{1}{m}\Upsilon n(\textbf{r},t)-\frac{\hbar^{2}}{2m^{2}}\frac{\triangle\sqrt{n(\textbf{r},t)}}{\sqrt{n(\textbf{r},t)}}+\frac{1}{m}V_{ext}(\textbf{r},t)=const.\end{equation}
Equation (\ref{ch int Koshy}) is the equation of the potential of velocity field evolution.

Having equations for the particle concentration and velocity field potential evolution we can derive equation for function $\Phi(\textbf{r},t)$, defined as
\begin{equation}\label{ch WF in medium}
\Phi(\textbf{r},t)=\sqrt{n(\textbf{r},t)}\exp\biggl(\frac{\imath}{\hbar}m\phi(\textbf{r},t)\biggr),\end{equation}
which called the macroscopic wave function, the order parameter or the wave function in the medium. This construction has macroscopical meaning due to it's definition via the macroscopic parameters.

Differentiating the macroscopic wave function (\ref{ch WF in medium}) with respect to time and using the continuity equation (\ref{ch continuity via vel}) and the  Cauchy-Lagrangian integral (\ref{ch int Koshy}) we find following non-linear Schrodinger equation
\begin{equation}\label{ch GP equation} \imath\hbar\partial_{t}\Phi(\textbf{r},t)=\biggl(-\frac{\hbar^{2}\nabla^{2}}{2m}+\mu(\textbf{r},t)+V_{ext}(\textbf{r},t)-\Upsilon\mid\Phi(\textbf{r},t)\mid^{2}\biggr)\Phi(\textbf{r},t),\end{equation}
which is well known  as the Gross-Pitaevskii equation ~\cite{Landau Vol 9}, ~\cite{L.P.Pitaevskii RMP 99}.
Operator $\hbar^{2}\nabla^{2}/2m$ arises in the Gross-Pitaevskii equation due to the quantum Bohm potential (\ref{ch Bohm tensor general}) in the Euler equation (\ref{ch eiler via vel and stress tensor}), or more precisely, it appears due to the approximate form of the quantum Bohm potential.

The wave function $\Phi(\textbf{r},t)$ is normalized by the condition
$$\int d\textbf{r}\Phi(\textbf{r},t)^{*}\Phi(\textbf{r},t)=N,$$
where $N$ is the number of particles in the system.

\section{Contribution of the temperature in dynamics of Bose particles in the first order by the interaction radius}

We describe QHD description of bosons at non-zero temperatures. Other method of the BEC description, for example the self-consistent Hartree-Fock-Bogoliubov approximation, can be found in Ref.s \cite{temp BEC}-\cite{Griffin 04}.

At low, but non-zero, temperatures part of Bose particles are in
the BEC state (on the ground energy level) and other part
distributes on excited states (non-condensed particles).
Traditionally one species of bosons in described conditions is
considered  as a mixture of two liquids. One liquid is the
particles in the BEC state, and another one is non-condensed
particles. Since we consider non ideal Bose gas, we have
interaction between particles, and in chosen model we have as
interaction between particles of each liquid and inter-liquid
interaction.

Interparticle interaction can lead to exchange of particles
between the two liquids (since these liquids contain atoms of the
same species), even if we keep system at fixed temperature.
Decreasing (increasing) of the system temperature leads to
decrease (increase) of a particles number in excited states, and,
correspondingly, it leads to increase (decrease) of a particles
number in the BEC state. In mentioned cases the number of
particles in each liquids changes, but total number of particles
does not change. Therefore we have the continuity equation for
total particles concentration $n$, which has usual form
\begin{equation}\label{ch temp cont B}\partial_{t}n(\textbf{r},t)+\nabla \textbf{j}=0,\end{equation}
where $\textbf{j}$ is the current of all particles, or the total
current, but for described above processes we have creation and
destruction particles, at transition between liquids, which cause
additional terms in the right-hand side of the continuity equation
corresponding to the partial concentration introduced for each
liquid.

We will consider a system of Bose particles at fixed non-zero
temperature and neglect by the particles transitions between the
liquids. In this case we the continuity equation for each liquid
in usual form
\begin{equation}\label{ch temp cont B}\partial_{t}n_{B}(\textbf{r},t)+\nabla(n_{B}(\textbf{r},t)\textbf{v}_{B}(\textbf{r},t))=0\end{equation}
for particles in the BEC state, and
\begin{equation}\label{ch temp cont n}\partial_{t}n_{n}(\textbf{r},t)+\nabla(n_{n}(\textbf{r},t)\textbf{v}_{n}(\textbf{r},t))=0.\end{equation}
for the non-condensed bosons, where the total particles
concentration $n(\textbf{r},t)$ and current
$\textbf{j}(\textbf{r},t)$ are divided on two parts
$$n(\textbf{r},t)=n_{B}(\textbf{r},t)+n_{n}(\textbf{r},t),$$
and
$$\textbf{j}(\textbf{r},t)=\textbf{j}_{B}(\textbf{r},t)+\textbf{j}_{n}(\textbf{r},t),$$
$n_{B}$, $n_{n}$ are particle concentrations for the BEC and
non-condensed particles, \emph{and} $\textbf{j}_{B}$ and
$\textbf{j}_{n}$ are corresponding currents, which have usual relation with the corresponding velocity fields $\textbf{j}_{B}=n_{B}\textbf{v}_{B}$ and
$\textbf{j}_{n}=n_{n}\textbf{v}_{n}$.

Next step is dividing of the momentum balance equation (the Euler equation) on two parts corresponding to condensed and non-condensed particle evolution.

We start with kinematic part of the momentum evolution which is given by the momentum current $\Pi^{\alpha\beta}$. It has bilinear structure on wave function as the concentration $n$ and particle current $\textbf{j}$. Thus we can represent the momentum current as the sum $\Pi^{\alpha\beta}=\Pi^{\alpha\beta}_{B}+\Pi^{\alpha\beta}_{n}$. Explicit form of $\Pi^{\alpha\beta}_{B}$ and $\Pi^{\alpha\beta}_{n}$ are given by formulas (\ref{ch Pi representation}) and (\ref{ch Bohm tensor general}), where we should put $n_{B}$, $v_{B}$ or $n_{n}$, $v_{n}$ instead of $n$ and $v$, \emph{and} neglect thermal pressure $p^{\alpha\beta}$.

For understanding of dynamical part of momentum evolution evolution of BEC
$\textbf{j}_{B}(\textbf{r},t)$ and non-condensed particles $\textbf{j}_{n}(\textbf{r},t)$
we need to consider formulas (\ref{ch second expansion of wave func}) and
(\ref{ch n2 long r}) in details.

The first multiplier in
formula (\ref{ch second expansion of wave func}), which has argument $(\textbf{r},t)$ is related to
the particle whose motion we consider. Other one particle wave
functions are related to the particles that influence on dynamic
of considered current. This is give us ability to obtain the
separate equation of dynamic atoms in the BEC state and the
non-condensed state.  If one-particle
wave function with argument $(\textbf{r},t)$ describe the BEC
state (has subindex "B"), we put this term in momentum balance
equation for the  BEC. In the case one-particle wave function with
argument $(\textbf{r},t)$ describe the non-condensed state we put
this term in the momentum balance equation for the non-condensed
particles.

If we consider dynamic of particle in the BEC state, it means a quantum state $f$ in formula (\ref{ch second expansion of wave func}) describes the BEC state we find that the first two terms in formula (\ref{ch n2 long r}) describes quantities related to the particles in different state, for our case it means that they describe interaction of condensed and non-condensed particles. Thus the square of concentration and the square of module of the density matrix lead to evolution of the BEC due to interaction with non-condensed particles, they trace arises as $2n_{n}n_{B}$. The last term in formula (\ref{ch second expansion of wave func}) describes two-particle in the same quantum state $f$. It gives us that the last term describes interaction between particles in the BEC state, and as we obtained in formula (\ref{ch corr.n2}) it equals to the square of the concentration of condensed particles.

Let's consider momentum evolution of non-condensed particles. Now a quantum state $f$ describes one of the non-ground states. We repeat that the first term (two first terms) in formula (\ref{ch second expansion of wave func}) (in formula (\ref{ch n2 long r})) describes quantities related to the particles in different quantum states. In the case of the BEC evolution we had fixed state of evaluating particles. Thus we had only one combination of of different states they are BEC state and some of non-condensed states. Now we have more possibilities. Considering particle in a non-condensed state we find that it can interact with a particle either in the BEC state or in other non-condensed state. Thus we obtain the sum of $2n_{n}n_{B}$ and $2n_{n}^{2}$. The last term gives additional contribution in non-condensed particles interaction. However, it was shown in Ref. \cite{Andreev arxiv ThPart} that it gives small contribution and we can neglect this term.

As a result we obtain
$$mn_{B}(\partial_{t}v^{\alpha}_{B}+v^{\beta}_{B}\nabla^{\beta}v^{\alpha}_{B})-\frac{\hbar^{2}}{4m}\partial^{\alpha}\triangle n_{B}$$
\begin{equation}\label{ch temp moment with temp with corr BEC}+\frac{\hbar^{2}}{4m}\partial^{\beta}\biggl(\frac{\partial^{\alpha}n_{B}\cdot\partial^{\beta}n_{B}}{n_{B}}\biggr)=-n_{B}\nabla^{\alpha}V_{ext}(\textbf{r},t)+\frac{1}{2}\Upsilon\partial^{\alpha}(2n_{B}n_{n}+n_{B}^{2}),\end{equation}
and
$$mn_{n}(\partial_{t}v^{\alpha}_{n}+v^{\beta}_{n}\nabla^{\beta}v^{\alpha}_{n})-\frac{\hbar^{2}}{4m}\partial^{\alpha}\triangle n_{n}$$
\begin{equation}\label{ch temp moment with temp with corr norm}+\frac{\hbar^{2}}{4m}\partial^{\beta}\biggl(\frac{\partial^{\alpha}n_{n}\cdot\partial^{\beta}n_{n}}{n_{n}}\biggr)=-n_{n}\nabla^{\alpha}V_{ext}(\textbf{r},t)+\frac{1}{2}\Upsilon\partial^{\alpha}(2n_{B}n_{n}+2n_{n}^{2}),\end{equation}
These equations have similar form, especially due to kinematic nature of the left-hand sides of the equations. The right-hand sides describe interaction with external field $V_{ext}$ and interparticle interaction proportional to the interaction constant $\Upsilon$. We have only one interaction constant since we consider one species in different quantum states. The right-hand side of equation (\ref{ch temp moment with temp with corr BEC}) contains two terms proportional to $\Upsilon$. One of this terms contains $n_{B}^{2}$ and describes interaction between particles in the BEC state. The term, containing $n_{B}n_{n}$, describes interaction between non-condensed particles and particles in the BEC state. Analogous terms in equation (\ref{ch temp moment with temp with corr norm}) have similar meaning, and term $n_{B}n_{n}$ present interaction between two fluids, and it's influence of non-condensed particles evolution. The last term in equation (\ref{ch temp moment with temp with corr norm}), proportional to $2n_{n}^{2}$, describes interaction in non-condensed particles.

The last term in equation (\ref{ch temp moment with temp with corr BEC}) emerges from the last term in formula (\ref{ch n2 long r}), and the last term in equation (\ref{ch temp moment with temp with corr norm}) appears from the two first terms in formula (\ref{ch n2 long r}), therefore it is quite normal that they have different coefficients.

\section{Quantum hydrodynamics of bosons at account of interaction up to the TOIR}

We have expanded derivation of both the Euler equation and corresponding non-linear Schrodinger equation, which is the Gross-Pitaevskii equation. In this section we will briefly discuss a contribution of the second and third terms in the quantum stress tensor.

In the second and third terms, as in the first one, the variables of the center of gravity and variables of the relative distance of
the particles can be separated. The second term in the stress tensor, for spherically symmetric interparticle potential, equals to zero due to the integral on the variable of the relative distance. Thus, we need to consider the third term only.

In the absence of particles in excited states, the quantum stress
tensor can be presented as
$$
\sigma^{\alpha\beta}_{BEC}(\textbf{r},t)=-\frac{1}{2}\Upsilon\delta^{\alpha\beta}\sum_{g}n_{g}(n_{g}-1)(|\varphi_{g}(\textbf{r},t)|^{2})^{2}$$
$$-\frac{1}{6}\Upsilon_{2}(\delta^{\alpha\beta}\triangle+2\partial^{\alpha}\partial^{\beta})\sum_{g}n_{g}(n_{g}-1)(|\varphi_{g}(\textbf{r},t)|^{2})^{2}$$
$$ -\Upsilon_{2}(\delta^{\alpha\beta}\delta^{\gamma\delta}+\delta^{\alpha\gamma}\delta^{\beta\delta}+\delta^{\alpha\delta}\delta^{\beta\gamma})\times $$
\begin{equation}\label{ch sigma in 2 or only BEC} \times\sum_{g}n_{g}(n_{g}-1)\Biggl(\varphi^{*}_{g}\varphi^{*}_{g}(\varphi_{g}\partial_{\gamma}\partial_{\delta}\varphi_{g}-\partial_{\gamma}\varphi_{g}\partial_{\delta}\varphi_{g})+h.c.\Biggr)
,\end{equation}
where we assumed that particles occupy a ground
quantum state described by wave function $\varphi_{g}$.

The quantum stress tensor in the FOIR approximation was found for
system of particles which are in some quantum state described with
wave function $\phi_{0}(\textbf{r},t)$ (see formula (\ref{ch
corr.n2})). It was managed to calculate the quantum stress tensor,
using intermediate formula (\ref{ch sigma in 2 or only BEC}), in
the TOIR approximation for approximately independent particles, as
for free particles and for particles in the parabolic trap
\cite{Andreev PRA08}. In this case the second term in formula
(\ref{ch sigma in 2 or only BEC}) simplifies and the last becomes equal to zero.

Method of calculation is the same as described above for the first term, but rather more complicated, so we do not present details.
Final form of the third term in the quantum stress tensor is
\begin{equation}\label{}\sigma^{\alpha\beta}(\textbf{r},t)=-\frac{1}{6}\Upsilon_{2}(\delta^{\alpha\beta}\triangle+2\partial^{\alpha}\partial^{\beta})n^{2}(\textbf{r},t),\end{equation}
where
\begin{equation}\label{ch Upsilon2}\Upsilon_{2}\equiv\frac{\pi}{30}\int
(r)^{5}\frac{\partial U(r)}{\partial r}dr\end{equation}
is the constant of interaction arising at account of the short-range interaction up to the TOIR, this definition differs from the definition presented in Ref. \cite{Andreev PRA08}, here we put factor $1/8$ in the integral (\ref{ch Upsilon2}).

Collecting parts of the quantum stress tensor arising in the FOIR and TOIR we find
\begin{equation}\label{ch sigma in 2 or boze pl w}\sigma^{\alpha\beta}(\textbf{r},t)=-\frac{1}{2}\Upsilon\delta^{\alpha\beta}n^{2}(\textbf{r},t)-\frac{1}{6}\Upsilon_{2}(\delta^{\alpha\beta}\triangle+2\partial^{\alpha}\partial^{\beta})n^{2}(\textbf{r},t),\end{equation}
which has form of an operator acting on the square of the particle concentration. Formula (\ref{ch sigma in 2 or boze pl w}) reveals symmetry of the quantum stress tensor: $\sigma^{\alpha\beta}=\sigma^{\beta\alpha}$, which exists in general formula (\ref{ch sigma shorted series}) and remains in approximate formula (\ref{ch sigma in 2 or boze pl w}).

Euler equation contains divergence of the quantum stress tensor which emerges as
\begin{equation}\label{ch sigma div in 2 or boze pl w}\partial^{\beta}\sigma^{\alpha\beta}(\textbf{r},t)=-\Upsilon n(\textbf{r},t)\partial^{\alpha}n(\textbf{r},t)-\frac{1}{2}\Upsilon_{2}\partial^{\alpha}\triangle n^{2}(\textbf{r},t).\end{equation}
We have found that the quantum stress tensor depends on higher than first spatial derivatives, see the second term in formula (\ref{ch sigma div in 2 or boze pl w}). Such dependence leads to nonlocal non-linear Schrodinger equation. Other nonlocal approximations for the BEC was considered in Ref.s \cite{Rosanov PL.A. 02}, \cite{Braaten PRA 01}.

At studying of quantum gases we have dial with boson-fermion and fermion-fermion mixtures. The QHD of ultracold fermions and boson-fermion mixtures was developed up to the TOIR approximation in Ref. \cite{Andreev PRA08}.

\section{Energy evolution up to TOIR approximation}

Energy evolution is associated with both the collective motion and
thermal motion of particles. We have no thermal motion in the BEC.
Thus, it was enough to use two hydrodynamical equations for the
BEC description. Non-condensed particles are involved in the
thermal motion. At low enough temperatures we can neglect by the
thermal pressure in the Euler equation for non-condensed particles.
Including thermal pressure contribution and influence of it's
evolution we have to have equation for the pressure evolution. In
most cases we can limit our treatment with consideration of the
scalar pressure which relates to the kinetic energy of the thermal
motion. We actually can derive the pressure evolution equation,
but it will be more complicated due to tensor nature of the
pressure, so we consider energy evolution only. We give definition
of whole energy, including energy of collective motion described
fully by the momentum density evolution, and we will extract
energy of thermal motion below.

Energy density for quantum system is defined as
\begin{equation}\label{ch en kin
def}\varepsilon(\textbf{r},t)=\int
dR\sum_{i=1}^{N}\delta(\textbf{r}-\textbf{r}_{i})\Biggl(\frac{1}{4m_{i}}\biggl(\psi^{*}\hat{\textbf{p}}^{2}_{i}\psi+c.c.\biggr)
+\frac{1}{2}\sum_{i, j\neq
i}^{N}U(|\textbf{r}_{i}-\textbf{r}_{j}|)\psi^{*}\psi\Biggr),\end{equation}
where the first term in the big brackets is the kinetic energy
density of $i$-th particle, and the second term is the density of
potential energy.

Differentiating the energy density (\ref{ch en kin def}) with respect to time we find general form of the energy evolution equation
\begin{equation}\label{ch eq bal en b}\partial_{t}\varepsilon(\textbf{r},t)+\nabla
\textbf{Q}(\textbf{r},t)=-\textbf{j}(\textbf{r},t)\nabla
V_{ext}(\textbf{r},t)+A(\textbf{r},t),\end{equation}
which has traditional structure and does not distinguish from classical case.

In equation (\ref{ch eq bal en b}) we have the energy flux $\textbf{Q}(\textbf{r},t)$ and the density of the work $A(\textbf{r},t)$. Due to our calculation these quantities arise in explicit form. Obtained equation is appropriate even for particles with long-range interparticle interaction, so explicit form of $\textbf{Q}(\textbf{r},t)$ and $A(\textbf{r},t)$ are correct in general case.

Let us present explicit form of the energy flux $\textbf{Q}(\textbf{r},t)$, which we separate on two parts having different meaning
\begin{equation}\label{pot en
  razl}Q^{\alpha}(\textbf{r},t)=Q^{\alpha}_{(kin)}(\textbf{r},t)+Q^{\alpha}_{(int)}(\textbf{r},t),
\end{equation}
the flux of the kinetic energy $Q^{\alpha}_{(kin)}(\textbf{r},t)$, and the flux of the potential energy $Q^{\alpha}_{(int)}(\textbf{r},t)$.

The flux of the kinetic energy has following representation via many-particle wave-function
\begin{equation}\label{pot en kin def b f}
Q^{\alpha}_{(kin)}(\textbf{r},t)=\int
dR\sum_{i=1}^{N_{b}}\delta(\textbf{r}-\textbf{r}_{i})\frac{1}{8m^{2}_{i}}
\Biggl((\hat{p}^{\alpha}_{i}\psi)^{*}\hat{\textbf{p}}^{2}_{i}\psi+\psi^{*}\hat{p}^{\alpha}_{i}\hat{\textbf{p}}^{2}_{i}\psi+c.c.\Biggr),
\end{equation}
the flux of the potential energy containing the interaction potential $U_{ij}$ arises as
\begin{equation}\label{ch pot en int def b f} Q^{\alpha}_{(int)}(\textbf{r},t)=\frac{1}{4}\int
dR\sum_{i,j\neq
i}^{N}\delta(\textbf{r}-\textbf{r}_{i})U_{ij}\frac{1}{m_{i}}\Biggl(\psi^{*}\hat{p}^{\alpha}_{i}\psi+(\hat{p}^{\alpha}_{i}\psi)^{*}\psi\Biggr),
\end{equation}
and the work density is
\begin{equation}\label{ch rabota b
b}A(\textbf{r},t)=-\frac{1}{2}\int dR\sum_{i,j\neq
i}^{N}\delta(\textbf{r}-\textbf{r}_{i})(\nabla^{\alpha}_{i}U_{ij}\Biggl(\frac{1}{2m_{i}}\psi^{*}\hat{p}^{\alpha}_{i}\psi+\frac{1}{2m_{j}}\psi^{*}\hat{p}^{\alpha}_{j}\psi+c.c.\Biggr).
\end{equation}

Using the fact that we consider short-range interaction, thus we
can use the method described in section (III) to represent the work
density and the potential energy flux in corresponding form. This
form appears to have very large form, so we present these quantity
separated on parts. We start with representation of the potential
energy flux via one-particle functions $\varphi_{g}$ describing
one-particle states, which can be occupied by particles. Dividing
$Q^{\alpha}_{(int)}$ on two parts
$$ Q^{\alpha}_{(int)}(\textbf{r},t)=Q^{\alpha}_{(int)d}(\textbf{r},t)+Q^{\alpha}_{(int)s}(\textbf{r},t). $$
Indexes $d$ and $s$ mean that particles under consideration are in different states and same state, correspondingly.

Explicit form of $Q^{\alpha}_{(int)d}$ arises as
$$Q^{\alpha}_{(int)d}=\frac{\hbar}{2\imath m}\Gamma_{1}\Biggl(\sum_{g,g',g\neq
  g'}n_{g}n_{g'}\varphi_{g}^{*}\varphi_{g'}^{*}(\partial^{\alpha}\varphi_{g})\varphi_{g'}-c.c.\Biggr)$$
$$+\frac{\hbar}{2\imath m}\Gamma_{2}\Biggl(2\sum_{g,g',g\neq
  g'}n_{g}n_{g'}\varphi_{g}^{*}\varphi_{g'}^{*}(\partial^{\alpha}\partial^{\beta}\partial^{\beta}\varphi_{g})\varphi_{g'}
-2\sum_{g,g',g\neq
  g'}n_{g}n_{g'}\varphi_{g}^{*}\varphi_{g'}^{*}(\partial^{\alpha}\partial^{\beta}\varphi_{g})\partial^{\beta}\varphi_{g'}$$
$$-\sum_{g,g',g\neq
  g'}n_{g}n_{g'}\Biggl((\partial^{\alpha}\varphi_{g}^{*})\varphi_{g'}^{*}(\partial^{\beta}\partial^{\beta}\varphi_{g})\varphi_{g'}
+\varphi_{g}^{*}(\partial^{\alpha}\varphi_{g'}^{*})(\partial^{\beta}\partial^{\beta}\varphi_{g})\varphi_{g'}\Biggr)$$
$$+2\sum_{g,g',g\neq
 g'}n_{g}n_{g'}\varphi_{g}^{*}(\partial^{\alpha}\varphi_{g'}^{*})(\partial^{\beta}\varphi_{g})\partial^{\beta}\varphi_{g'}-c.c.\Biggr)$$
\begin{equation}\label{ch Q d} +\frac{\hbar}{2\imath m}\Gamma_{2}\partial^{\beta}\partial^{\beta} \Biggl(\sum_{g,g',g\neq
 g'}n_{g}n_{g'}\varphi_{g}^{*}\varphi_{g'}^{*}(\partial^{\alpha}\varphi_{g})\varphi_{g'}-c.c.\Biggr)\end{equation}
here and in the following formula for $Q^{\alpha}_{(int)s}$ we use following designations
\begin{equation}\label{ch Gamma1}
\Gamma_{1}=\int d\textbf{r}U(r),\end{equation}
and
\begin{equation}\label{ch Gamma2}\Gamma^{\alpha\beta}_{2}=\int r^{\beta}r^{\gamma}U(r)d\textbf{r}=\delta^{\alpha\beta}\Gamma_{2},\end{equation}
where
\begin{equation}\label{ch Gamma2 b}\Gamma_{2}\equiv\frac{4\pi}{3}\int U(r)r^{4} dr  .\end{equation}
Coefficient $\Gamma_{1}$ describing interaction in the FOIR approximation is the Gross-Pitaevskii interaction constant $g$, which is simply related with $\Upsilon=-g=-\Gamma$. $\Gamma_{2}$ arises in terms coming out in the TOIR approximation, so it has to be connected with the $\Upsilon_{2}$. Integrating definition of $\Upsilon_{2}$ (\ref{ch Upsilon2}) by parts we find that $\Upsilon_{2}=-\Gamma_{2}$.

The potential energy flux related to interaction of particles in the same quantum state has form of
$$Q^{\alpha}_{(int)s}(\textbf{r},t)=\frac{\hbar}{4\imath m}\Gamma_{1}\Biggl(\sum_{g}n_{g}(n_{g}-1)\varphi_{g}^{*}\varphi_{g}^{*}\varphi_{g}\partial^{\alpha}\varphi_{g}-c.c.\Biggr)$$
$$+\frac{\hbar}{2\imath m}\Gamma_{2}\Biggl(\sum_{g}n_{g}(n_{g}-1)\varphi_{g}^{*}\varphi_{g}^{*}\varphi_{g}\partial^{\alpha}\partial^{\beta}\partial^{\beta}\varphi_{g}(\textbf{r},t)
-\sum_{g}n_{g}(n_{g}-1)\varphi_{g}^{*}\varphi_{g}^{*}(\partial^{\beta}\varphi_{g})\partial^{\alpha}\partial^{\beta}\varphi_{g}$$
$$-\sum_{g}n_{g}(n_{g}-1)(\partial^{\alpha}\varphi_{g}^{*})\varphi_{g}^{*}\varphi_{g}\partial^{\beta}\partial^{\beta}\varphi_{g}
+\sum_{g}n_{g}(n_{g}-1)\varphi_{g}^{*}(\partial^{\alpha}\varphi_{g}^{*})(\partial^{\beta}\varphi_{g})\partial^{\beta}\varphi_{g}-c.c.\Biggr)$$
\begin{equation}\label{ch Q s} +\frac{\hbar}{4\imath m}\Gamma_{2}\partial^{\beta}\partial^{\beta} \Biggl(\sum_{g}n_{g}(n_{g}-1)\varphi_{g}^{*}(\textbf{r},t)\varphi_{g}^{*}\varphi_{g}\partial^{\alpha}\varphi_{g}-c.c.\Biggr). \end{equation}

We are interested in studying of particle dynamics at low temperatures, consequently we suppose that the ground energy state and an interval of low energy states are macroscopically occupied. We also can approximately neglect tail  of states with larger energies which are 	
faintly occupied. Therefore we will consider cases then $n_{g}(n_{g}-1)\approx n_{g}^{2}$ in $\textbf{Q}_{int (s)}$ and $A_{s}$.

We have explicitly presented intermediate formulas for the potential energy flux up to the TOIR approximation. Now we consider analogous formulas for the work.

We divide the work on two parts following the recipe as for potential energy flow
$$ A(\textbf{r},t)=A_{d}(\textbf{r},t)+A_{s}(\textbf{r},t),$$
$A_{s}$ ($A_{d}$) is the work related to two in different (same) quantum states.

Following formula for the work $A_{d}$ emerges from our calculations
$$A_{d}(\textbf{r},t)=\frac{\hbar}{2\imath
m}\Upsilon\partial_{\alpha}\Biggl(\sum_{g,g',g\neq
g'}n_{g}n_{g'}\varphi^{*}_{g}\varphi^{*}_{g'}\varphi_{g'}\partial^{\alpha}\varphi_{g}-c.c.\Biggr)$$
$$ +\frac{\hbar}{3!\imath m}\Upsilon_{2}(\delta^{\alpha\beta}\delta^{\gamma\delta}+\delta^{\alpha\gamma}\delta^{\beta\delta}+\delta^{\alpha\delta}\delta^{\beta\gamma}) \partial^{\beta}\partial^{\gamma}\partial^{\delta}\Biggl(\sum_{g,g',g\neq
g'}n_{g}n_{g'}\varphi^{*}_{g}\varphi^{*}_{g'}\varphi_{g'}\partial^{\alpha}\varphi_{g}-c.c.\Biggr)$$
$$ +\frac{\hbar}{2\imath m}\Upsilon_{2}(\delta^{\alpha\beta}\delta^{\gamma\delta}+\delta^{\alpha\gamma}\delta^{\beta\delta}+\delta^{\alpha\delta}\delta^{\beta\gamma})
\partial^{\beta} \Biggl(\sum_{g,g',g\neq
g'}n_{g}n_{g'}\varphi^{*}_{g}\varphi^{*}_{g'}\varphi_{g'}\partial^{\alpha}\partial^{\gamma}\partial^{\delta}\varphi_{g}$$
$$+\sum_{g,g',g\neq
g'}n_{g}n_{g'}\varphi^{*}_{g}\varphi^{*}_{g'}(\partial^{\gamma}\partial^{\delta}\varphi_{g})\partial^{\alpha}\varphi_{g'}
-2\sum_{g,g',g\neq g'}n_{g}n_{g'}\varphi^{*}_{g}\varphi^{*}_{g'}(\partial^{\delta}\varphi_{g'})\partial^{\alpha}\partial^{\gamma}\varphi_{g}$$
$$-\sum_{g,g',g\neq g'}n_{g}n_{g'}\biggl((\partial^{\alpha}\varphi^{*}_{g})\varphi^{*}_{g'}\varphi_{g'}\partial^{\gamma}\partial^{\delta}\varphi_{g}
+(\partial^{\alpha}\varphi^{*}_{g'})\varphi^{*}_{g}\varphi_{g'}\partial^{\gamma}\partial^{\delta}\varphi_{g}\biggr)$$
\begin{equation}\label{ch A d}+\sum_{g,g',g\neq g'}n_{g}n_{g'}\biggl((\partial^{\alpha}\varphi^{*}_{g})\varphi^{*}_{g'}(\partial^{\gamma}\varphi_{g'})\partial^{\delta}\varphi_{g}
+(\partial^{\alpha}\varphi^{*}_{g'})\varphi^{*}_{g}(\partial^{\gamma}\varphi_{g'})\partial^{\delta}\varphi_{g}\biggr)-c.c.\Biggr).\end{equation}

In terms proportional $\Upsilon_{2}$ we meet the tensor structure $\delta^{\alpha\beta}\delta^{\gamma\delta}+\delta^{\alpha\gamma}\delta^{\beta\delta}+\delta^{\alpha\delta}\delta^{\beta\gamma}$, which has obtained in the quantum stress tensor $\sigma^{\alpha\beta}$ (\ref{ch sigma in 2 or only BEC}), in the terms which appeared in the TOIR approximation. It relates to the fact that the second interaction constant $\Upsilon_{2}$ emerges as a fourth rank tensor, and the structure under discussion reflects symmetries of this tensor. We also meet this structure in following formula for $A_{s}$, which is
$$A_{s}(\textbf{r},t)=\frac{\hbar}{4\imath
m}\Upsilon\partial_{\alpha}\Biggl(\sum_{g}n_{g}(n_{g}-1)\varphi^{*}_{g}\varphi^{*}_{g}\varphi_{g}\partial^{\alpha}\varphi_{g}
-c.c.\Biggr)$$
$$+\frac{1}{2}\frac{\hbar}{3!\imath m}\Upsilon_{2}(\delta^{\alpha\beta}\delta^{\gamma\delta}+\delta^{\alpha\gamma}\delta^{\beta\delta}+\delta^{\alpha\delta}\delta^{\beta\gamma})
\partial^{\beta}\partial^{\gamma}\partial^{\delta}\Biggl(\sum_{g}n_{g}(n_{g}-1)\varphi^{*}_{g}\varphi^{*}_{g}\varphi_{g}\partial^{\alpha}\varphi_{g}-c.c.\Biggr)$$
$$+\frac{\hbar}{4\imath m}\Upsilon_{2}(\delta^{\alpha\beta}\delta^{\gamma\delta}+\delta^{\alpha\gamma}\delta^{\beta\delta}+\delta^{\alpha\delta}\delta^{\beta\gamma})
\partial^{\beta}\Biggl(\sum_{g}n_{g}(n_{g}-1)\varphi^{*}_{g}\varphi^{*}_{g}\varphi_{g}\partial^{\alpha}\partial^{\gamma}\partial^{\delta}\varphi_{g}$$
$$+\sum_{g}n_{g}(n_{g}-1)\varphi^{*}_{g}\varphi^{*}_{g}(\partial^{\gamma}\partial^{\delta}\varphi_{g})\partial^{\alpha}\varphi_{g} -2\sum_{g}n_{g}(n_{g}-1)\varphi^{*}_{g}\varphi^{*}_{g}(\partial^{\delta}\varphi_{g})\partial^{\alpha}\partial^{\gamma}\varphi_{g}$$
\begin{equation}\label{ch A s}-2\sum_{g}n_{g}(n_{g}-1)\varphi^{*}_{g}(\partial^{\alpha}\varphi^{*}_{g})\varphi_{g}\partial^{\gamma}\partial^{\delta}\varphi_{g}+2\sum_{g}n_{g}(n_{g}-1)\varphi^{*}_{g}(\partial^{\alpha}\varphi^{*}_{g})(\partial^{\gamma}\varphi_{g})\partial^{\delta}\varphi_{g}
-c.c.\Biggr).\end{equation}

These expressions (\ref{ch Q d}), (\ref{ch Q s}), (\ref{ch A d}) and (\ref{ch A s}) can be used for developing of different approximations for the quantum gases evolution, and we use them for one simplest approximation of approximately free particles. Thus, we use plane waves
\begin{equation}\label{ch pl.wave view}\varphi_{p}(\textbf{r},t)=\frac{1}{\sqrt{V}}\exp\biggl(-\frac{\imath}{\hbar}(\varepsilon_{p}t-\textbf{p}\textbf{r})\biggr),\end{equation}
for description of one-particle states.

To find closed description for the energy evolution we need to
present the particles current $j^{\alpha}$, momentum current
$\Pi^{\alpha\beta}$
\begin{equation}\label{ch j varphi} j^{\alpha}(\textbf{r},t)=\frac{1}{2m}\sum_{g}n_{g}\biggl(\varphi_{g}^{*}(\textbf{r},t)\hat{p}^{\alpha}\varphi_{g}(\textbf{r},t)+c.c.\biggr) ,\end{equation}
\begin{equation}\label{ch Pi varphi}
\Pi^{\alpha\beta}(\textbf{r},t)=\frac{1}{4m}\sum_{g}n_{g}\biggl(\varphi_{g}^{*}(\textbf{r},t)\hat{p}^{\alpha}\hat{p}^{\beta}\varphi_{g}(\textbf{r},t)+(\hat{p}^{\alpha
*}\varphi_{g}^{*}(\textbf{r},t))\hat{p}^{\beta}\varphi_{g}(\textbf{r},t)+c.c.\biggr),
\end{equation}

\begin{equation}\label{ch varepsilon varphi}\varepsilon=\frac{1}{4m}\sum_{g}n_{g}\biggl(\varphi_{g}^{*}(\textbf{r},t)\hat{p}^{2}\varphi_{g}(\textbf{r},t)+c.c.\biggr), \end{equation}
and
\begin{equation}\label{ch Q varphi}Q^{\alpha}=\frac{1}{8m}\sum_{g}n_{g}\biggl((\hat{p}^{2}\varphi_{g})^{*}(\textbf{r},t)\hat{p}^{\alpha}\varphi_{g}(\textbf{r},t)+\varphi_{g}^{*}(\textbf{r},t)\hat{p}^{\alpha}\hat{p}^{2}\varphi_{g}(\textbf{r},t)+c.c.\biggr). \end{equation}
These quantities get following form for the plane wave function (\ref{ch pl.wave view}) $n=\Sigma_{\textbf{p}}n_{\textbf{p}}/V$, $j^{\alpha}=\Sigma_{\textbf{p}}p^{\alpha}n_{\textbf{p}}/V$, $\varepsilon_{kin}=\Sigma_{\textbf{p}}\textbf{p}^{2}n_{\textbf{p}}/V$, $\Pi^{\alpha\beta}=\Sigma_{\textbf{p}}p^{\alpha}p^{\beta}n_{\textbf{p}}/V$, and $Q^{\alpha}=\Sigma_{\textbf{p}}p^{\alpha}\textbf{p}^{2}n_{\textbf{p}}/V$.
These formulas allow us obtain following formulas for the potential energy flux and the work
\begin{equation}Q_{(int) d}^{\alpha}=-\Upsilon n j^{\alpha}-\frac{3}{4}\Upsilon_{2}\triangle (n j^{\alpha})-\frac{2\Upsilon_{2}}{\hbar^{2}}(m\varepsilon_{(kin)}j^{\alpha}-nQ^{\alpha}),\end{equation}
\begin{equation}Q_{(int) s}(\textbf{r},t)=-\frac{1}{2m}\Upsilon \Lambda^{\alpha}(\textbf{r},t)-\frac{1}{2m}\Upsilon_{2}\triangle\Lambda^{\alpha}(\textbf{r},t),\end{equation}
\begin{equation}A_{d}=\Upsilon\partial^{\alpha}(nj^{\alpha})+\Upsilon_{2}\partial^{\alpha}\triangle(nj^{\alpha})+\frac{4m\Upsilon_{2}}{\hbar^{2}}\partial^{\alpha}(j^{\alpha}\varepsilon_{kin}+j^{\beta}\Pi^{\alpha\beta}-3nQ^{\alpha}_{kin}),\end{equation}
and
\begin{equation}A_{s}(\textbf{r},t)=\frac{1}{2m}\Upsilon\partial^{\alpha}\Lambda^{\alpha}(\textbf{r},t)+\frac{1}{2m}\partial^{\alpha}\triangle\Lambda^{\alpha}(\textbf{r},t),\end{equation}
where
\begin{equation}\Lambda^{\alpha}(\textbf{r},t)=\frac{1}{V}\sum_{\textbf{p}}p^{\alpha}n_{\textbf{p}}^{2}.\end{equation}

Fixing local macroscopic quantum state via knowledge of macroscopic quantum parameters $n$, $\textbf{j}$, $\varepsilon$ ..., we limit a set of microscopic quantum states, which needs for description of particle system. Overwise, if we know microscopic quantum state of a system (we know in which quantum state each particle is)  we uniquely obtain macroscopic parameters describing system.

\section{Three particle interaction between particles in the BEC}

QHD of the BEC with three-particle interaction was developed in Ref. \cite{Andreev arxiv ThPart}. Before presenting of the basic equation let us briefly discuss meaning of the three-particle interaction (TPI). We can present interaction between three particles $V_{3}(\textbf{r}_{i}, \textbf{r}_{j}, \textbf{r}_{k})$ in the form
$$V_{3}(\textbf{r}_{i}, \textbf{r}_{j},
\textbf{r}_{k})=U(\textbf{r}_{ij})$$
\begin{equation}\label{ch tp three part int} +U(\textbf{r}_{ik})+U(\textbf{r}_{jk})+U(\textbf{r}_{ij},\textbf{r}_{ik},\textbf{r}_{jk}),\end{equation}
where $\textbf{r}_{ij}=\textbf{r}_{i}-\textbf{r}_{j}$, $U_{ij}=U(\mid \textbf{r}_{ij}\mid)$ is the binary interaction potential, $U_{ijk}=U(\mid
\textbf{r}_{ij}\mid,\mid \textbf{r}_{ik}\mid,\mid\textbf{r}_{kj}\mid)$ is the TPI potential that does not contain combination of binary potentials $U_{ij}$. It means that if we have deal with the simultaneously collision of three particles we can expect that the two-particle potential is enough for description of this process. If particle has no inner structure, as electron, for example, they have potential of two-particle interaction only. If particles have an inner structure, as atoms and molecules, we can expect that the TPI $U_{ijk}$ can give contribution in the collision.

The Hamiltonian of the system under consideration has the form:
$$\hat{H}=\sum_{i}\frac{1}{2m_{i}}\hat{p}^{\alpha}_{i}\hat{p}^{\alpha}_{i}+\sum_{i}V_{ext}(\textbf{r}_{i},t)$$
\begin{equation}\label{ch tp ugam ht}+\frac{1}{2}\sum_{i,j\neq i}U_{ij}+\frac{1}{6}\sum_{i,j\neq i;k\neq i,j}U_{ijk},\end{equation}
where $\hat{p}^{\alpha}_{i}=-\imath\hbar\nabla_{i}$ is the momentum operator of the i-th particle, $m_{i}$-is the mass of the i-th particle. Let us consider the TPI nonequivalent to combination of the binary interaction described by $U_{ij}$. Interaction of three or more particles at the same time by means of binary potential is described by $U_{ij}$ and have no connection with the TPI.

$$m\partial_{t}j^{\alpha}(\textbf{r},t)+\partial_{\beta
}\Pi^{\alpha\beta}(\textbf{r},t)=-\int
d\textbf{r}'(\nabla^{\alpha}U(\textbf{r},\textbf{r}'))n_{2}(\textbf{r},\textbf{r}',t)$$
\begin{equation}\label{ch tp Ebi}-\int d\textbf{r}'\int d\textbf{r}''(\partial^{\alpha}U(\textbf{r},\textbf{r}',\textbf{r}''))n_{3}(\textbf{r},\textbf{r}',\textbf{r}'',t)-n(\textbf{r},t)\nabla^{\alpha}V_{ext}(\textbf{r},t).\end{equation}
we have introduced the three-particle concentration
$$n_{3}(\textbf{r},\textbf{r}',\textbf{r}'',t)=\int dR\sum_{i,j\neq i;k\neq i,j}\delta(\textbf{r}-\textbf{r}_{i})$$
\begin{equation} \label{ch tp n3def}\times\delta(\textbf{r}'-\textbf{r}_{j})\delta(\textbf{r}''-\textbf{r}_{k})\psi^{+}(R,t)\psi(R,t).\end{equation}

It is necessary to consider variables
of the center of gravity and variables of the relative distance of the particles for three particles
$$\begin{array}{ccc} \textbf{R}_{ijk}=\frac{1}{3}(\textbf{r}_{i}+\textbf{r}_{j}+\textbf{r}_{k}) ,& \textbf{r}_{ij}=\textbf{r}_{i}-\textbf{r}_{j},&\textbf{r}_{ik}=\textbf{r}_{i}-\textbf{r}_{k}  \end{array},$$
\begin{equation} \label{ch tp def}\textbf{r}_{jk}=\textbf{r}_{j}-\textbf{r}_{k}=\textbf{r}_{ik}-\textbf{r}_{ij}.\end{equation}
Using these variables we can represent the force field in terms of the quantum stress tensor
\begin{equation}\label{ch tp eq b j}\partial_{t}j^{\alpha}(\textbf{r},t)+\frac{1}{m}\partial_{\beta
 }(\Pi^{\alpha\beta}(\textbf{r},t)+\sigma^{\alpha\beta}(\textbf{r},t))=-\frac{1}{m}n(\textbf{r},t)\nabla^{\alpha}V_{ext}(\textbf{r})
.\end{equation}

Limiting our consideration by the FOIR approximation for two- and three-particle interaction we find following representation for the quantum stress tensor
$$ \sigma^{\alpha\beta}(\textbf{r},t)=-\frac{1}{2}\int
dR\sum_{i,j.i\neq j}\delta(\textbf{r}-\textbf{R}_{ij})\frac{r^{\alpha}_{ij}r^{\beta}_{ij}}{\mid\textbf{r}_{ij}\mid}\frac{\partial U(\textbf{r}_{ij})}{\partial\mid\textbf{r}_{ij}\mid}\psi^{+}\psi$$
$$$$
\begin{equation} \label{ch tp sigma in 1 or}-\frac{1}{9}\int dR\sum_{i;i\neq j;k\neq i,j}\Biggl((r_{ij}^{\beta}\partial_{j}^{\alpha}+r_{ik}^{\beta}\partial_{k}^{\alpha})U(r_{ij},r_{ik},\mid\textbf{r}_{ij}-\textbf{r}_{ik}\mid)\Biggr)\delta(\textbf{r}-\textbf{R}_{ijk})\psi^{+}\psi,\end{equation}
where
$$\psi(R,t)=\psi(...,R_{ijk},...,R_{ijk},...,t)$$
for $\psi$ function in the term
describing the binary interaction, similarly in the term describing the
three-particle interaction the psi-function has following structure
$$\psi(R,t)=\psi(...,R_{ijk},...,R_{ijk},...,R_{ijk},...,t).$$

Separating variables
of the center of gravity and variables of the relative distance of the particles in the quantum stress tensor we obtain
$$
\sigma^{\alpha\beta}(\textbf{r},t)=-\frac{1}{2}Tr(n_{2}(\textbf{r},\textbf{r}',t))\int
d \textbf{r}\frac{r^{\alpha}r^{\beta}}{r}\frac{\partial
U(r)}{\partial r}$$
$$-\frac{1}{9}Tr (n_{3}(\textbf{r}
,\textbf{r}',\textbf{r}'',t))\times$$
\begin{widetext}
\begin{equation} \label{ch tp sigma in 1 or hz n2}\times\int
d\textbf{r}_{12}d\textbf{r}_{13}\Biggl(r_{12}^{\beta}\frac{\partial}{\partial
r_{2}^{\alpha}}U(r_{12},r_{13},\mid\textbf{r}_{12}-\textbf{r}_{13}\mid)+r_{13}^{\beta}\frac{\partial}{\partial
r_{3}^{\alpha}}U(r_{12},r_{23},\mid\textbf{r}_{12}-\textbf{r}_{13}\mid)\Biggr),\end{equation}
\end{widetext}
where
$$ Tr f(\textbf{r},\textbf{r}')=f(\textbf{r},\textbf{r}) ,$$
and
$$Tr f(\textbf{r},\textbf{r}',\textbf{r}'')=f(\textbf{r},\textbf{r},\textbf{r}).$$

Calculation of the three-particle concentration requires further
expansion of the many-particle wave function in comparison with
formula (\ref{ch second expansion of wave func}) \cite{Andreev
arxiv ThPart}.

For particles to be found in
the Bose condensation state contribution of $Tr n_{3}$ reduces to
\begin{equation}\label{ch temp corr.n3}\widetilde{m}_{B}(\textbf{r},t)=\sum_{g}n_{g}(n_{g}-1)(n_{g}-2) |\varphi_{g}(\textbf{r},t)|^{6}=n_{B}^{3}(\textbf{r},t),\end{equation}
which was calculated analogously to (\ref{ch corr.n2}).

In the issue we get the Euler equation
$$m\partial_{t}v^{\alpha}(\textbf{r},t)+\frac{1}{2}m\partial^{\alpha}v^{2}(\textbf{r},t)+m\partial^{\alpha}\mu(\textbf{r},t)$$
$$-\frac{\hbar^{2}}{4m}\partial^{\alpha}\triangle n+\frac{\hbar^{2}}{4m}\partial^{\beta}\biggl(\frac{\partial^{\alpha}n\cdot\partial^{\beta}n}{n}\biggr)-\Upsilon\partial^{\alpha}n(\textbf{r},t)$$
\begin{equation}\label{ch tp eiler gen}-2\chi^{\alpha\beta} n^{2}(\textbf{r},t)\partial^{\beta}n(\textbf{r},t)=-
n(\textbf{r},t)\nabla^{\alpha}V_{ext}(\textbf{r},t).
\end{equation}
where the interaction constant for the TPI arises in the tensor
form
$$\chi^{\alpha\beta}\equiv-\frac{1}{6}\int d\textbf{r}_{1}d\textbf{r}_{2}\Biggl(\Biggl(\frac{r_{1}^{\alpha}r_{1}^{\beta}}{r_{1}}\partial_{1}+\frac{r_{2}^{\alpha}r_{2}^{\beta}}{r_{2}}\partial_{2}+2\frac{(r_{1}^{\alpha}-r_{2}^{\alpha})(r_{1}^{\beta}-r_{2}^{\beta})}{\mid \textbf{r}_{1}-\textbf{r}_{2}\mid}\partial_{3}\Biggr)$$
\begin{equation}\label{tp chi}\times U(r_{1},r_{2},\sqrt{r_{1}^{2}+r_{2}^{2}+2r_{1}r_{2}\cos\Omega})  \Biggr),\end{equation}
where $\Omega$ is the angle between  $\textbf{r}_{1}$ and
$\textbf{r}_{2}$, $r_{1}$ and $r_{2}$ are modules of the vectors
$\textbf{r}_{1}$ and $\textbf{r}_{2}$, \emph{and} $\partial_{1}$,
$\partial_{2}$, $\partial_{3}$ are derivatives of function $U$ on
its arguments $r_{1}$, $ r_{2}$, $
\sqrt{r_{1}^{2}+r_{2}^{2}+2r_{1}r_{2}\cos\Omega}$ correspondingly.
We can see that $\chi^{\alpha\beta}=\chi^{\beta\alpha}$.

Reducing our description to the scalar TPI we can rewrite the
Euler equation as
$$m\partial_{t}v^{\alpha}(\textbf{r},t)+\frac{1}{2}m\partial^{\alpha}v^{2}(\textbf{r},t)+m\partial^{\alpha}\mu(\textbf{r},t)$$
$$-\frac{\hbar^{2}}{4m}\partial^{\alpha}\triangle n+\frac{\hbar^{2}}{4m}\partial^{\beta}\biggl(\frac{\partial^{\alpha}n\cdot\partial^{\beta}n}{n}\biggr)-\Upsilon\partial^{\alpha}n(\textbf{r},t)$$
\begin{equation}\label{ch tp eiler TPI scal}-\chi\partial^{\alpha}n^{2}(\textbf{r},t)=-\partial^{\alpha}V_{ext}(\textbf{r},t),\end{equation}
where scalar TPI constant has following form
$$\chi\equiv-\frac{1}{3}\int d\textbf{r}_{1}d\textbf{r}_{2}\Biggl(\Biggl(r_{1}\partial_{r_{1}}+\mid\textbf{r}_{1}-\textbf{r}_{2}\mid\partial_{3}\Biggr)$$
\begin{equation}\label{ch tp chi isotrop}\times U(r_{1},r_{2},\sqrt{r_{1}^{2}+r_{2}^{2}+2r_{1}r_{2}\cos\Omega})  \Biggr).\end{equation}

Including of the three-particle interaction in the FOIR approximation gives the one additional term in the momentum balance equation and it also gives contribution in the non-linear Schrodinger equation, the method of derivation described above for Bose system with the two-particle interaction, which reveals as a nonlinearities of fifth degree
\begin{equation}\label{ch tp u GP} \imath\hbar\partial_{t}\Phi(\textbf{r},t)=\Biggl(-\frac{\hbar^{2}\nabla^{2}}{2m}+\mu(\textbf{r},t)+V_{ext}(\textbf{r},t)-\Upsilon\mid\Phi(\textbf{r},t)\mid^{2}-\chi\mid\Phi(\textbf{r},t)\mid^{4}\Biggr)\Phi(\textbf{r},t)
.\end{equation}
That is well known Gross-Pitaevskii equation
~\cite{L.P.Pitaevskii RMP 99},
with the nonlinearity
of the fifth degree ~\cite{Kovalev FNT 76,Barashenkov PLA
88}, which arises due to the TPI, in approximation of the scalar TPI.

\section{Dispersion of linear collective excitations in BEC}

In this chapter we have considered the two-particle interaction up to the TOIR and the three-particle interaction in the FOIR. In this section we consider contribution of these interactions in the dispersion of the collective excitations of the BEC. Let's write the set of the QHD equations including described interactions, which is
\begin{equation}\label{ch continuity via vel for disp}
\partial_{t}n+\nabla(n\textbf{v})=0,\end{equation}
and
$$mn\partial_{t}v^{\alpha}+\frac{1}{2}mn\nabla^{\alpha}v^{2}-\frac{\hbar^{2}}{4m}\partial^{\alpha}\triangle n+\frac{\hbar^{2}}{4m}\partial^{\beta}\biggl(\frac{\partial^{\alpha}n\cdot\partial^{\beta}n}{n}\biggr)$$
\begin{equation}\label{ch eiler boze TOIR and TPI} -\frac{1}{2}\Upsilon\partial_{\alpha
 }n^{2}-\chi\partial^{\alpha}n^{3}-\frac{1}{2}\Upsilon_{2}\partial_{\alpha}\triangle n^{2}=-n\nabla^{\alpha}V_{ext}.\end{equation}
The last three terms in the left-hand side of equation (\ref{ch eiler boze TOIR and TPI}) describe interparticle interaction, which are two-particle interaction in the FOIR approximation, three-particle interaction in the FOIR approximation, and two-particle interaction in the TOIR approximation, correspondingly.

We consider the small perturbation of equilibrium state like
\begin{equation}\label{ch tp equlib state BEC}\begin{array}{ccc}n=n_{0}+\delta n,& v^{\alpha}=0+v^{\alpha},&\end{array}\end{equation}
Substituting these relations into system of equations  (\ref{ch continuity via vel for disp}) and (\ref{ch eiler boze TOIR and TPI}) and neglecting nonlinear
terms, we obtain a system of linear homogeneous equations in
partial derivatives with constant coefficients. Passing to the
following representation for small perturbations $\delta f$
\begin{equation}\label{ch tp FFF}\delta f =f(\omega, \textbf{k}) exp(-\imath\omega t+\imath \textbf{k}\textbf{r}) \end{equation}
yields the homogeneous system of algebraic equations. The
magnitude of concentration of the BEC is assumed to have a nonzero
value. Expressing all the quantities entering the system of
equations in terms of the concentration of BEC, we come to the
dispersion equation for elementary excitations
\begin{equation}\label{ch tp dispersion General} \omega^{2}=\Biggl(\frac{\hbar^{2}}{4m^{2}}+\frac{n_{0}\Upsilon_{2}}{m}\Biggr)k^{4}-\Biggl(\frac{\Upsilon n_{0}}{m}+\frac{2\chi}{m}n_{0}^{2}\Biggr)k^{2} .\end{equation}
The first and third terms corresponds to the Bogoliubov spectrum, which also appears from the Gross-Pitaevskii equation. The first term arises from the linear part of the quantum Bohm potential. The third term exists due to the two-particle short range interaction in the first order by the interaction radius. Account of the two-particle interaction up to the TOIR gives the second term, which leads to dependence of coefficient at $k^{4}$ on the equilibrium concentration $n_{0}$. The last term is caused by the TPI.

Nonlinear shift of the frequency of collective excitations was calculated up to the TOIR approximation \cite{Andreev Izv.Vuzov. 09 1}, change of the form of the bright soliton in the BEC due to account of the short-range interaction up to the TOIR was obtained in Ref. \cite{Andreev RPJ 11}. Moreover, it was shown that considering of the short-range interaction up to the TOIR gives new solutions \cite{Andreev MPL B 12}, namely new soliton solution was found. This soliton exists in the repulsing BEC, and it reveals as area of compression, so it was called bright-like soliton, when the bright soliton exists in the attractive BEC. At usual conditions the repulsing BEC reveals the dark soliton, which is an area of rarefication.

\section{Quantum hydrodynamics of the dipolar BEC}

One of the more excited topics in BEC studies is the electrically polarized BEC (EPBEC), which is the BEC of polar molecules. Quantum gases of particles having electric dipole moment add two interesting fundamental properties. These are long-range interaction and anisotropy of the interparticle interaction.

The Gross-Pitaevskii equation with the cubic nonlinearities has been used for studying of the unpolarized BEC with two-particle interaction. Taking into account the three-particle interaction we come to the non-linear Schrodinger equation containing the quintic nonlinearity along with the cubic nonlinearity. Thus, we have generalization of the Gross-Pitaevskii equation. The cubic and quantic nonlinearities play role of the potential energy of interparticle interaction in the non-linear Schrodinger equation. Therefore, if we want to generalized the Gross-Pitaevskii equation for description of the EPBEC we need to add the potential energy of dipole-dipole interaction. In this way was obtained the generalized Gross-Pitaevskii equation for the EPBEC \cite{Goral PRA 00}-\cite{Fischer PRA 06R}
$$\imath\hbar\partial_{t}\Phi(\textbf{r},t)=\biggl(-\frac{\hbar^{2}\nabla^{2}}{2m}+\mu(\textbf{r},t)+V_{ext}(\textbf{r},t)+g\mid\Phi(\textbf{r},t)\mid^{2}$$
\begin{equation}\label{ch di BEC GP eq for introduction}+d^{2}\int d\textbf{r}'
\frac{1-3\cos^{2}\theta'}{|\textbf{r}-\textbf{r}'|^{3}}\mid\Phi(\textbf{r}',t)\mid^{2}\biggr)\Phi(\textbf{r},t)
.\end{equation}
In equation (\ref{ch di BEC GP eq for introduction}) following designation are used: $\Phi(\textbf{r},t)$ is the macroscopic wave function, $\mu$ is the chemical potential, $V_{ext}$ is the potential of external field, $g$ is the constant of short-range interaction, $d$ is the dipole electric moment of single atom, $m$ is the mass of particles and $\hbar$ is the Planck constant divided by $2\pi$. The last term in equation (\ref{ch di BEC GP eq for introduction}) describes the dipole-dipole interaction.

We derive the QHD equations for the EPBEC \cite{Andreev PRB 11}, \cite{Andreev arxiv 12 02}, and \cite{Andreev RPJ 12} from the many-particle Schrodinger equation with the following Hamiltonian
$$\hat{H}=\sum_{i}\Biggl(\frac{1}{2m_{i}}\hat{\textbf{p}}_{i}^{2}-d_{i}^{\alpha}E_{i,ext}^{\alpha}+V_{trap}(\textbf{r}_{i},t)\Biggr)$$
\begin{equation}\label{ch di BEC Hamiltonian}+\frac{1}{2}\sum_{i,j\neq i}\Biggl(U_{ij}-d_{i}^{\alpha}d_{j}^{\beta}G_{ij}^{\alpha\beta}\Biggr).\end{equation}
The first term here is the operator for kinetic energy. The second
term represents the interaction between the dipole moment
$d_{i}^{\alpha}$ and the external electrical field. The subsequent
terms represent short-range $U_{ij}$ and dipole-dipole
interactions between particles, respectively. The Green's function
for dipole-dipole interaction is taken as
$G_{ij}^{\alpha\beta}=\nabla^{\alpha}_{i}\nabla^{\beta}_{i}(1/r_{ij})$.

Usually people use following Hamiltonian of dipole-dipole interaction
\begin{equation}\label{ch di BEC pot of dd int} H_{dd}=\frac{\delta^{\alpha\beta}-3r^{\alpha}r^{\beta}/r^{2}}{r^{3}}d_{1}^{\alpha}d_{2}^{\beta},\end{equation}
which is enough for mechanical description of several particles motion, but it does not enough for construction of the field theory. Thus, in this paper for interaction of electric dipoles we use the following, more general, Hamiltonian:
$$H_{dd}=-\partial^{\alpha}\partial^{\beta}\frac{1}{r}\cdot d_{1}^{\alpha}d_{2}^{\beta}.$$
There is well-known identity
\begin{equation}\label{ch di BEC togdestvo}-\partial^{\alpha}\partial^{\beta}\frac{1}{r}= \frac{\delta^{\alpha\beta}-3r^{\alpha}r^{\beta}/r^{2}}{r^{3}}+\frac{4\pi}{3}\delta^{\alpha\beta}\delta(\textbf{r}),\end{equation}
so we can see the difference between usually using Hamiltonian and one's used in this paper. The corrections of our selection followed from the fact that the equations obtained in the paper coincide to the Maxwell equations. At derivation of the QHD equation in self-consistent approximation lead to field equation
\begin{equation}\label{ch di BEC field good}\nabla\textbf{E}(\textbf{r},t)=-4\pi \nabla\textbf{P}(\textbf{r},t),\end{equation}
but if we used Hamiltonian (\ref{ch di BEC pot of dd int}) we  would obtain
\begin{equation}\label{ch di BEC field wrong}\nabla\textbf{E}(\textbf{r},t)=\frac{8\pi}{3} \nabla\textbf{P}(\textbf{r},t),\end{equation}
instead of (\ref{ch di BEC field good}).

Electric field arising in the QHD equations in the self-consistent field approximation at using of the Hamiltonian (\ref{ch di BEC pot of dd int}) satisfy to the two Maxwell equation (\ref{ch di BEC field good}) and
\begin{equation}\label{ch di BEC field curl}\nabla\times \textbf{E}=0.\end{equation}

Scheme of derivation of the QHD equations is the same as described in previous sections. Thus, we have following set of equations, which contains additional equations in compare with the unpolarized systems described below. The continuity equation
\begin{equation}\label{ch di BEC cont eq}\partial_{t}n+\partial^{\alpha}(nv^{\alpha})=0\end{equation}
has the same form as usual.

The momentum balance equation for the polarized BEC has the form
$$mn(\partial_{t}+\textbf{v}\nabla)v^{\alpha}-\frac{\hbar^{2}}{4m}\partial^{\alpha}\triangle
n+\frac{\hbar^{2}}{4m}\partial^{\beta}\Biggl(\frac{\partial^{\alpha}n\cdot\partial^{\beta}n}{n}\Biggr)$$
\begin{equation}\label{ch di BEC bal imp eq short}=\Upsilon n\partial^{\alpha}n+P^{\beta}\partial^{\alpha}E^{\beta},
\end{equation}
and contains one additional term. New quantity has appeared in the momentum balance equation, which is the electric dipole moment density
\begin{equation}\label{ch def polar density}\textbf{P}(\textbf{r},t)=\int dR\sum_{i}\delta(\textbf{r}-\textbf{r}_{i})\textbf{d}_{i}\psi^{*}(R,t)\psi(R,t).\end{equation}
It is obviously very important quantity for description of polarized BEC evolution, so that we should consider their evolution and influence of the polarization evolution on dynamics of hydrodynamic variables (particle concentration $n$ and velocity field $\textbf{v}$). When we consider system of fully polarized dipoles at zero temperature we can rewrite polarization $\textbf{P}=\textbf{d}n$. At assumption that particles dynamic do not lead to change of the dipoles direction we can conclude that changing of polarization caused by changing of particles position, so polarization increase in area where concentration increase and vice-versa. The generalized Gross-Pitaevskii equation (\ref{ch di BEC GP eq for introduction}) corresponds to described case. Generally speaking we should include dipoles direction evolution, and consequently more detailed polarization evolution. Even in ferromagnetics, where magnetic moments are kept parallel due to strong exchange interaction, there are waves of dipoles direction (spin waves). Therefore, we need to derive equation of the polarization evolution.

Considering two kind of interaction, the short-range interaction and dipole-dipole interaction, the last one is an example of the long-range interaction, we have dial with different approximations for each interaction. Approximations for short-range interaction were minutely considered in the beginning of the chapter. We have used the self-consistent field approximation \cite{Andreev PRB 11} for dipole-dipole interaction. Briefly, this approximation corresponds to approximate representation of two-particle function as a product of corresponding one particle function. More precisely, we can explain it on example of two-particle concentration. The self-consistent field approximation corresponds to the first term in formula (\ref{ch n2 long r}), which is the general representation of the two-particle concentration in terms of one-particle states $\varphi_{g}$.

As usual we differentiate the polarization (\ref{ch def polar density}) with respect to time and using Schrodinger equation with the Hamiltonian (\ref{ch di BEC Hamiltonian}) the equation of polarization evolution appears as
\begin{equation}\label{ch di BEC eq polarization}\partial_{t}P^{\alpha}(\textbf{r},t)+\partial^{\beta}R^{\alpha\beta}(\textbf{r},t)=0,\end{equation}
$R^{\alpha\beta}(\textbf{r},t)$ is the current of polarization.

Using a self-consistent field approximation of the dipole-dipole interaction we obtain an equation for the polarization current $R^{\alpha\beta}(\textbf{r},t)$ evolution
$$\partial_{t}R^{\alpha\beta}+\partial^{\gamma}\biggl(R^{\alpha\beta}v^{\gamma}+R^{\alpha\gamma}v^{\beta}-P^{\alpha}v^{\beta}v^{\gamma}\biggr)-\frac{\hbar^{2}}{4m^{2}}\partial_{\beta}\triangle P^{\alpha}$$
\begin{equation}\label{ch di BEC eq for pol current gen selfconsist
appr}+\frac{\hbar^{2}}{8m^{2}}\partial^{\gamma}\biggl(\frac{\partial_{\beta}P^{\alpha}\partial_{\gamma}n}{n}+\frac{\partial_{\gamma}P^{\alpha}\partial_{\beta}n}{n}\biggr)=\frac{1}{m}\Upsilon\partial^{\beta}\biggl(nP^{\alpha}\biggr)+\frac{\sigma}{m}\frac{P^{\alpha}P^{\gamma}}{n}\partial^{\beta}E^{\gamma},\end{equation}
where the second term, containing three terms in large brackets, in the left-hand side is the convective part of the polarization current evolution. The third and fourth terms describe an analog of the quantum Bohm potential. We have two terms in the right-hand side describe influence of interaction on the polarization current evolution. The first term in the right-hand side presents contribution of the short-range interaction, where we have $\Upsilon$ as the signature. The very last term of the formula (\ref{ch di BEC eq for
pol current gen selfconsist appr}) describes interaction of dipoles with electric field and includes both external
electrical field and a self-consistent field that
created by particle dipoles. This term contain numerical constant $\sigma$.

Considering monochromatic collective excitations in the EPBEC
\begin{equation}\label{ch dip FFF}\delta f =f(\omega, \textbf{k}) exp(-\imath\omega t+\imath \textbf{k}\textbf{r}) \end{equation}
choosing equilibrium condition as $n=n_{0}$, $\textbf{v}=0$, $\textbf{P}=P_{0}\textbf{e}_{z}$, and $R^{\alpha\beta}=0$, including that equilibrium external electric field is $\textbf{E}=E_{0}\textbf{e}_{z}$,
we find that the dispersion dependence for the collective excitation in the EPBEC
can be expressed in the form of
$$\omega^{2}=\frac{1}{2m}\Biggl(\frac{\hbar^{2}k^{4}}{2m}+ 4\pi\sigma\frac{P_{0}^{2}k^{2}}{n_{0}}-\frac{3}{2}\Upsilon n_{0}k^{2}$$
\begin{equation}\label{ch di BEC general disp dep}\pm\sqrt{\biggl(\frac{1}{2}\Upsilon n_{0}k^{2}+ 4\pi\sigma\frac{P_{0}^{2}k^{2}}{n_{0}}\biggr)^{2}-8\pi\Upsilon k^{4}P_{0}^{2}}\Biggr),\end{equation}
quantities with subindex $0$ are constants.
This formula shows that in the EPBEC exists two waves, due to two signs in front of the square root, instead of one wave existing in an unpolarized BEC (\ref{ch tp dispersion General}) considered in previous section.

Comparing obtained dispersion dependence for EPBEC (\ref{ch di BEC general disp dep}) with the one derived in the absence of the electric dipole moment (\ref{ch tp dispersion General}) for $\Upsilon_{2}=0$ and $\chi=0$, since we have not considered the TPI and two-particle interaction in the TOIR approximation, we derive $P_{0}\rightarrow 0$ limit of formula (\ref{ch di BEC general disp dep}), which appears as
\begin{equation}\label{ch di BEC P0 1}\omega^{2}=\frac{1}{m}\biggl(\frac{\hbar^{2}k^{4}}{4m}-\Upsilon n_{0}k^{2}+\frac{8\pi P_{0}^{2}k^{2}}{n_{0}}\biggr),\end{equation}
for minus in front of the square root in formula (\ref{ch di BEC general disp dep}) and
\begin{equation}\label{ch di BEC P0 2}\omega^{2}=\frac{1}{m}\biggl(\frac{\hbar^{2}k^{4}}{4m}-\frac{1}{2}\Upsilon n_{0}k^{2}+\frac{8\pi (\sigma-1)P_{0}^{2}k^{2}}{n_{0}}\biggr)\end{equation}
for plus  in front of the square root. Tracking dependence on $\Upsilon$ we get that formula (\ref{ch di BEC P0 1}) corresponds to the polarizationless solution (\ref{ch tp dispersion General}).

Dipole molecules have been used for reaching of EPBEC has large electric dipole moment \cite{Ni PCCP 09}. Making estimation we find that in this, most interesting case, terms proportional to $P_{0}$ play leading role. Or more precisely, we have that $4\pi P_{0}^{2}/n_{0}\gg\Upsilon n_{0}$ and $4\pi P_{0}^{2}/n_{0}\gg\hbar^{2}k^{2}/2m $. As consequence we can simplify formula (\ref{ch di BEC general disp dep}) and find
\begin{equation}\label{ch di BEC P big 1}\omega^{2}=\frac{1}{m}\biggl(\frac{\hbar^{2}k^{4}}{4m}+\frac{4-5\sigma}{\sigma}\Upsilon n_{0}k^{2}\biggr),\end{equation}
for minus in front of the square root in formula (\ref{ch di BEC general disp dep}) and
\begin{equation}\label{ch di BEC P big 2}\omega^{2}=\frac{1}{m}\biggl(\frac{\hbar^{2}k^{4}}{4m}-\frac{\sigma+4}{\sigma}\Upsilon n_{0}k^{2}+\frac{4\pi P_{0}^{2}k^{2}}{n_{0}}\biggr).\end{equation}
for plus  in front of the square root.

Dipole-dipole interaction is the anisotropic interaction, so we can expect anisotropy of the collective excitation spectrum, especially including the fact that considered equilibrium condition is also anisotropic -- equilibrium electric field gives preferential direction. Nevertheless, dispersion dependence (\ref{ch di BEC general disp dep}) is isotropic. It happens due to consideration of the longitudinal waves, since we have used electrostatic Maxwell equations (\ref{ch di BEC field good}) and (\ref{ch di BEC field curl}). Contribution of the transverse electric field in the wave propagation was considered in Ref. \cite{Andreev arxiv 12 transv dip BEC}, where full set of the Maxwell equation along with the described in this section set of QHD equation. It was shown that dispersion dependence becomes anisotropic, and presented here formula (\ref{ch di BEC general disp dep}) transforms very little with replacement $P_{0}$ on $P_{0}\cos\theta$, where $\theta$ is the angle between direction of equilibrium electric field and direction of wave propagation.

\section{Conclusion}

We have described main results obtained at development of the QHD method for quantum gases. Main point of the QHD method is equation derivation from many-particle Schrodinger equation, which explicitly models microscopic quantum particle motion. Modeling particle system we include the main properties of particles and corresponding inter-particle interactions. We have obtained QHD equations for the BEC with two- and three-particle short range interaction. We derived the quantum stress tensor, which detailed analysis allows us to construct QHD equations. For two-particle interaction we have considered the quantum stress tensor up to the third order by the interaction radius, and for the TPI we have limited our consideration by the FOIR. We have specially considered long-range interaction between electrical dipole moments of neutral particles being in the BEC state. In this case we have had to derive equations for polarization evolution in addition to the continuity and the Euler equation, which make up usual QHD. By means of the QHD method we consider influence of non-zero temperature on evolution of bosons deriving two-fluid model and considering energy balance equation for non-condensed bosons. We have obtained contribution of described interactions dispersion of collective waves in the BEC.


\begin{thebibliography}{17}

\bibitem{Shirkov UFN 09} D. V. Shirkov,
Phys. Usp.
\textbf{52} 549 (2009).


\bibitem{Landau Vol 9} L. D. Landau and E. M. Lifshitz, \textit{Statistical Physics, Part 2:
Theory of the Condensed State}, Course of theoretical Physics
Vol. 9 (Pergamon Press, London, 1987).

\bibitem{L.P.Pitaevskii RMP 99} F. Dalfovo, S. Giorgini, L. P. Pitaevskii,
and S. Stringari, Rev. Mod. Phys. \textbf{71}, 463 (1999).



\bibitem{Kovalev FNT 76} A. S. Kovalev and A. M. Kosevich, Fiz.
Nizk. Temp. \textbf{2}, 913 (1976).

\bibitem{Barashenkov PLA 88} I. V. Barashenkov and V. G. Makhankov, Phys.
Lett. A. \textbf{128}, 52 (1988).

\bibitem{Abdullaev PRA 01} F. Kh. Abdullaev,
 A. Gammal, Lauro Tomio, and T. Frederico,
  Phys. Rev. A. \textbf{63} 043604 (2001).

\bibitem{Bedaque PRL 00} P. F. Bedaque, E. Braaten, and H.-W.
Hammer,  Phys. Rev. Lett. \textbf{85}, 908 (2000).

\bibitem{Jack
PRL 02} M. W. Jack,  Phys. Rev. Lett. \textbf{89}, 140402 (2002).



\bibitem{Goral PRA 00} K. Goral, K. Rzazewski, and T.
Pfau, Phys. Rev. A \textbf{61}, 051601(R) (2000).

\bibitem{Santos PRL 00} L. Santos, G.V. Shlyapnikov, P. Zoller, and M.
Lewenstein, Phys. Rev. Lett. \textbf{85}, 1791 (2000).

\bibitem{Yi PRA 00} S. Yi and L. You, Phys. Rev. A, \textbf{61}, 041604(R) (2000).

\bibitem{Ticknor PRL 11}  C. Ticknor, R. M. Wilson, and J. L.
Bohn, Phys. Rev. Lett. \textbf{106}, 065301 (2011).

\bibitem{Fischer PRA 06R} Uwe R.
Fischer, Phys. Rev. A \textbf{73}, 031602(R) (2006).

\bibitem{MaksimovTMP 1999} L. S. Kuz'menkov and S. G. Maksimov,  Theoretical and Mathematical Physics \textbf{118} 227 (1999).


\bibitem{MaksimovTMP 2001} L. S. Kuz'menkov, S. G. Maksimov, and V. V. Fedoseev, Theoretical and Mathematical
Physics, \textbf{126} 110 (2001).

\bibitem{Marklund PRL 07} M. Marklund and G. Brodin, Phys. Rev. Lett. \textbf{98}, 025001 (2007).

\bibitem{Andreev PRB 11} P. A. Andreev, L. S. Kuz'menkov, M. I. Trukhanova, Phys. Rev. B \textbf{84}, 245401 (2011).


\bibitem{Andreev PRA08} P. A. Andreev, L. S. Kuz'menkov, Phys. Rev. A \textbf{78}, 053624
(2008).


\bibitem{Landau Vol 3} E. M. Lifshitz and E. M. Lifshitz, \emph{Quantum Mechanics}
(Heinemann, Oxford, 1999).



\bibitem{Klimontovich book} Yu. L. Klimontovich, \emph{Statistical Physics} [in Russian], Nauka, Moscow (1982); English transl., Harwood, New
York (1986).


\bibitem{Weinberg book} S. Weinberg, \emph{Gravitation and Cosmology} (John Wiley and Sons, Inc., New York, 1972).


\bibitem{Landau 6} L. D. Landau and E. M. Lifshitz, \emph{Hydrodynamics} (Science, Moscow,
1986).


\bibitem{Shveber}  S. Schweber, \emph{An
Introdution to Relativistic Quantum Field Theory} (Evantson,
peterson; New York, Elmsford, 1961).



\bibitem{Rosanov PL.A. 02} N. N. Rosanov, A. G. Vladimirov, D. V.
Skryabin, W. J. Firth, Phys. Lett. A. \textbf{293}, 45 (2002).

\bibitem{Braaten PRA 01} E. Braaten, H.-W. Hammer, and Shawn Hermans,
Phys. Rev. A. \textbf{63}, 063609 (2001).


\bibitem{Andreev arxiv ThPart} P. A. Andreev, arXiv:1109.0896.

\bibitem{Andreev Izv.Vuzov. 09 1} P. A. Andreev, L. S. Kuz'menkov,
 Russian Physics Journal \textbf{52}, 912 (2009).

\bibitem{Andreev RPJ 11} P. A. Andreev and M. I. Trukhanova, Russian Physics
Journal \textbf{53}, 912 (2011).



\bibitem{Andreev MPL B 12} P. A. Andreev, L. S.
Kuz'menkov, Mod. Phys. Lett. B \textbf{26}, 1250152 (2012).






\bibitem{temp BEC} A. Griffin, Phys. Rev. B \textbf{53}, 9341 (1996).


\bibitem{Griffin arXiv 2011} E. Arahata, T. Nikuni, A. Griffin, Phys. Rev. A \textbf{84}, 053612 (2011).

\bibitem{Griffin 98 4044} T. Nikuni, A. Griffin, Phys. Rev. A \textbf{58}, 4044 (1998).

\bibitem{Griffin 97} A. Griffin, E. Zaremba, Phys. Rev. A \textbf{56}, 4839 (1997).

\bibitem{Griffin 98 4695} E. Zaremba, A. Griffin and T. Nikuni, Phys. Rev. A \textbf{57}, 4695 (1998).

\bibitem{Griffin 04} T. Nikuni, A. Griffin, Phys. Rev. A \textbf{69}, 023604 (2004).





\bibitem{Andreev arxiv 12 02} P. A. Andreev and L. S. Kuz'menkov,
arXiv:1201.2440.

\bibitem{Andreev RPJ 12} P. A. Andreev, Russian Physics
Journal \textbf{54}, 1360 (2012).


\bibitem{Ni PCCP 09} K.-K. Ni, S. Ospelkaus, D. J. Nesbitt, J. Ye and D. S. Jin, Phys. Chem. Chem. Phys. \textbf{11}, 9626 (2009).

\bibitem{Andreev arxiv 12 transv dip BEC} P. A.  Andreev and L. S. Kuz'menkov,
arXiv:1208.1000.

%
\end{thebibliography}
\end{document}